\documentclass[pra,aps,twocolumn,preprintnumbers,amsmath,amssymb, superscriptaddress, showkeys,nofootinbib]{revtex4}
\usepackage{amsmath,amsfonts,amstext,txfonts,amssymb,graphics,rotating,
dcolumn,graphics,array,multirow,mathrsfs,gensymb,graphicx,units,braket,xfrac}

\makeatletter
\let\l@addto@macro\relax
\makeatother
\usepackage[fontsize=10pt]{scrextend}

\usepackage[version=3]{mhchem} 
\usepackage[T1]{fontenc} 
\usepackage{natbib}

\begin{document}
\title{Quantum redirection of antenna absorption to photosynthetic reaction centres}
\author{Felipe Caycedo-Soler}
\affiliation{Institute of Theoretical Physics and Integrated Quantum Science and Technology IQST, University of Ulm, Albert-Einstein-Allee 11, D - 89069 Ulm, Germany}
\author{Christopher A Schroeder}
\affiliation{Institute of Theoretical Physics and Integrated Quantum Science and Technology IQST, University of Ulm, Albert-Einstein-Allee 11, D - 89069 Ulm, Germany}
\altaffiliation{Joint Quantum Institute, Department of Physics, University of Maryland and National Institute of Standards and Technology, College Park, MD 20742, USA}
\author{Caroline Autenrieth }
\affiliation{Department of Bioenergetics, Institute of Biomaterials and Biomolecular Systems, University of Stuttgart, Pfaffenwaldring 57, D - 70569 Stuttgart, Germany}
\author{Arne Pick}
\affiliation{Institute of Theoretical Physics and Integrated Quantum Science and Technology IQST, University of Ulm, Albert-Einstein-Allee 11, D - 89069 Ulm, Germany}
\author{Robin Ghosh }
\affiliation{Department of Bioenergetics, Institute of Biomaterials and Biomolecular Systems, University of Stuttgart, Pfaffenwaldring 57, D - 70569 Stuttgart, Germany}
\author{Susana F. Huelga}
\affiliation{Institute of Theoretical Physics and Integrated Quantum Science and Technology IQST, University of Ulm, Albert-Einstein-Allee 11, D - 89069 Ulm, Germany}
\author{Martin B. Plenio}
\affiliation{Institute of Theoretical Physics and Integrated Quantum Science and Technology IQST, University of Ulm, Albert-Einstein-Allee 11, D - 89069 Ulm, Germany}


\begin{abstract}
The early steps of photosynthesis involve the photo-excitation of  reaction centres (RCs)  and light-harvesting (LH) units. Here, we show that the --historically overlooked--   excitonic delocalisation across RC and LH pigments  results in a redistribution of dipole strengths that benefits the absorption cross section of the optical bands associated with the RC of several species. While we prove that this redistribution is  robust to the microscopic details of the dephasing between these units in the purple bacterium  {\it Rhodospirillum rubrum}, we are able to show that the redistribution witnesses  a more fragile, but persistent, coherent population dynamics which directs excitations from the LH {\it towards} the RC units under incoherent illumination and  physiological conditions. Stochastic  optimisation allows us to delineate clear guidelines and develop  simple analytic expressions,  in order to achieve directed coherent population dynamics in artificial nano-structures. 
\end{abstract}
\keywords{fick}

\maketitle

Photosynthesis -- the conversion of sunlight to chemical energy -- is fundamental for supporting life on our planet. Despite its importance, the physical principles that underpin the primary steps of photosynthesis, from photon absorption, to excitonic dynamics and electronic charge separation, remain to be understood in full. Excitonic delocalisation between tightly-packed pigments, such as within the RC {\it or} within the LH units, has been recognised to be of considerable importance for characterising their individual optical responses and for determining the associated time-scales for excitation energy transfer steps \cite{Jordanides_2001JPCB,Renger2006,VanGrondelle_JP2006,Timpmann_2005CPL,silbey_PRL_2006,OlayaLF+08}. 
More recently, the study of coherent effects in these biologically relevant systems has  attracted increasing attention due to the observation of long-lasting oscillatory signals measured with optical time-resolved techniques \cite{Engel_Nature2006,EngelNJP,ColliniWW+10,PanitchayangkoonVA+10,Hildner2013,Romero_NPhys2014,Ogilvie_NChem2014}. These results  have driven a wave of theoretical work aimed at understanding the microscopic mechanisms that may underpin persistent coherent signals\cite{Jonas_PNAS2012,Plenio_JCP2013,Caycedo_2012,Christensson_JPCB2012,KolliRS+12,ChinPR+13,HuelgaP13} and encouraged the discussion of the significance of coherent dynamics for efficient energy transfer \cite{Chin2010,Mohseni08,PlenioH08,Womick_2011,Strumpfer_JCP2012}.  These studies, though,  have discussed less on the impact of the coherent  RC-LH dynamics on optical spectra, likely biased by the observation that energy migration between RC and LH units is mainly driven by incoherent excitonic transfer \cite{Strumpfer_JCP2012}. 

As we will show here, the theoretical examination of the rapid time-scale inherent of absorption spectra underlines its importance as a useful tool to encode in the amplitudes -and not in their shifts- of these spectra, the effects of the moderate coupling between RC and LH pigments, despite the aggressive dephasing environment intrinsic to these photosynthetic complexes. Across several species, we find a noticeable  increase of amplitude of bands associated to RC transitions when they interact with the LH units, with respect to these bands from the isolated RC pigments, in a phenomenon we term absorption redistribution to the RC. The detailed study of this absorption redistribution, permits to understand a more subtle phenomenon present under physiological conditions, which we refer to as population redirection, and that represents an   increase in the RC population  driven by coherent dynamics, albeit is triggered by incoherent illumination. Although this redirection  in photosynthetic structures is relatively small with respect to the population  driven by the subsequent incoherent dynamics, we are able to show how it can be largely amplified in artificial devices, by following the guidelines observed from natural structures.

\begin{figure}
\centering
\includegraphics[width=70mm]{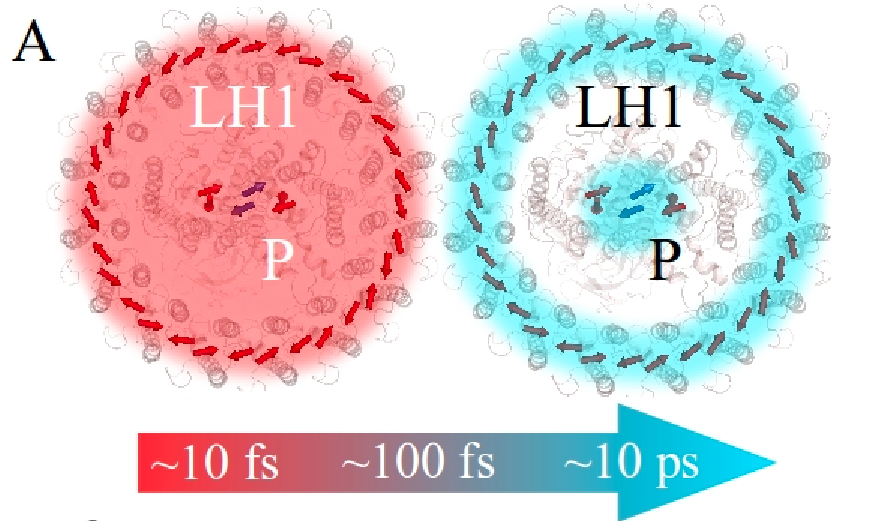}\\
\includegraphics[width=80mm]{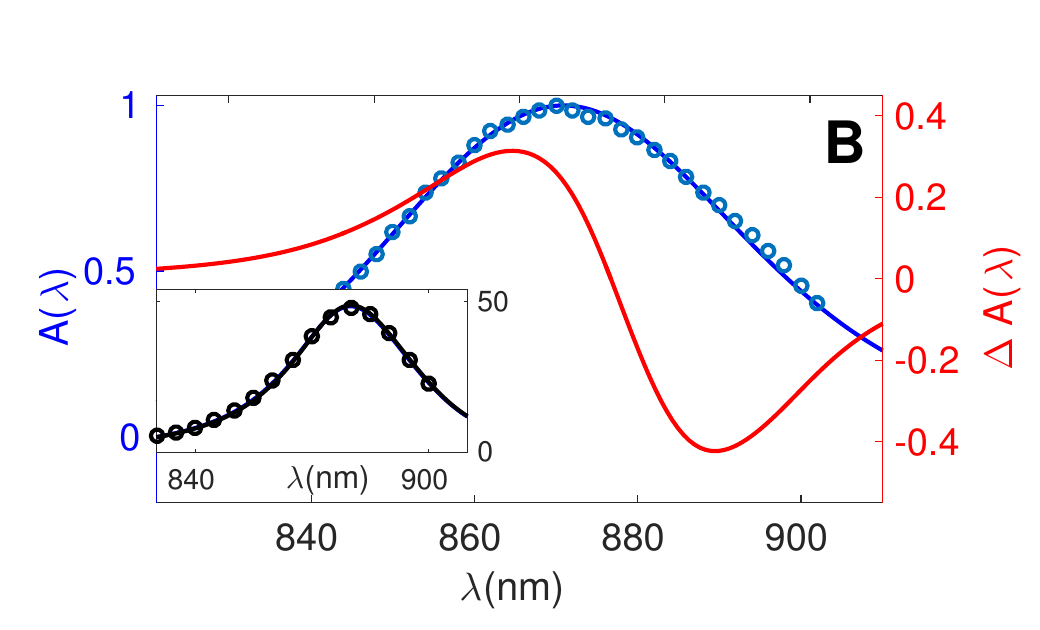}\\
\includegraphics[width=35mm]{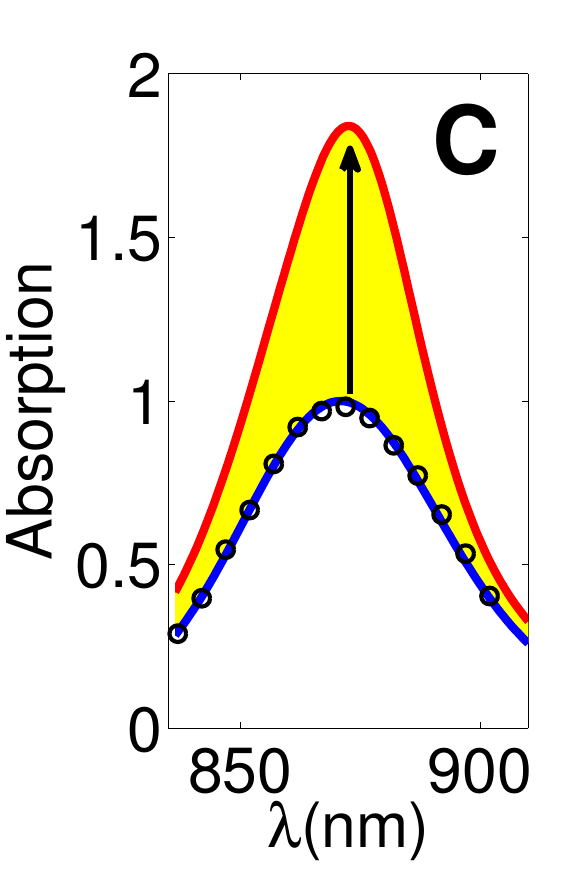}
\includegraphics[width=35mm]{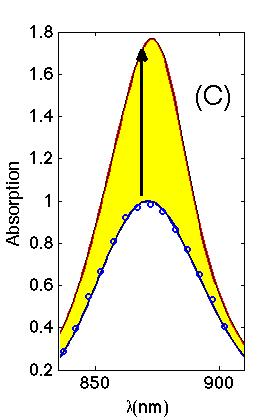}

\caption{ {\bf Temporal evolution of coherence and spectral signatures of full core complex delocalisation} {\bf A} Schematic representation of the {\it R. rubrum} RC-LH1 BChl pigments (see SI I.A), with excitons that within a few tens to hundred femtoseconds are delocalised over the entire core complex, before environment-induced dephasing  forces excitations to reside {\it either} on the antenna {\em or} the RC (blue shading).  {\bf B } Spectra from isolated RC  (blue theory, circles experiment). In red is presented the difference between the calculated  RC-like   and the isolated RC spectra (red). The inset in {\bf B} shows  the calculated (continuous) and experimental (circles) spectra of isolated LH1.  In {\bf C} RC (blue) and RC-like (red) spectra are compared. In {\bf D} the LH1 (blue) and LH1-like (red) spectra are shown.  In {\bf C} and {\bf D} yellow areas and arrows highlight the magnitude and the direction of  the absorption redistribution due to the RC-LH1 coherent coupling. All spectra are normalised to the maximum height of the P870 peak.}\label{fig0}
\end{figure}

\paragraph*{The principles of absorption redistribution}
The fundamental principle that motivates this work is illustrated in Fig.~1A. Our model system for  exemplifying the role of RC-LH coherent dynamics, is the core complex of the purple bacterium {\it R. rubrum}\cite{Ghosh}, composed of an LH (LH1) ring of $N_{LH}=32$ bacteriochlorophyll pigments (BChl) and a ``special pair'' ($P$, with $N_{RC}=2$ pigments) of BChl molecules which mediate the primary process of light-induced charge separation. The absorption spectrum $A(\omega=2\pi c/\lambda)$ of the $Q_y$ transition of the $P$ pigments peaks around $\lambda=870$ nm (circles in Figs.~1B and D), whereas the LH1 exhibits a single absorption band at $\lambda=880$ nm (inset in Fig.~1B and  Fig.~1D), denoted the P870 and B880 bands, respectively.  The single LH1 band arises mainly from a doublet of states which are bright due to the pigments' circular arrangement  with the individual pigments' $Q_y$ transition dipoles $\vec{d}_i$ arranged almost tangentially to the LH1 circumference.  Due to the fact that recently published spectra  confirmed that purple bacteria ecosystems of peat lakes and costal waters are dominated by red and near-IR light \cite{Stomp_2007ISME},  the near-IR $Q_y$ transition is the most relevant in {\it R. rubrum} under physiological conditions. As is illustrated in this figure, several time-scales affect the dynamics of excitations in the core complex. In the very early stages of these dynamics, the coherent interaction between LH and RC pigments results in excitonic delocalisation extended over RC and LH1 pigments. Thereby, observables able to probe such a short time-scale will, in principle, lead to specific signatures that may allow to determine the impact of the short lived coherent RC-LH exchange.  

The calculation of the absorption spectrum, $A(\omega)=\int_0^\infty\mbox{Tr}\{ D e^{{\cal U} t} D\rho_{ss}\} e^{i\omega t}dt$ depends on the  system's stationary state  $\rho_{ss}$, in order to obtain the dipole-dipole correlation function (DDCF), DDCF$(t)=\mbox{Tr}\{ D e^{{\cal U} t}D\rho_{ss}\}$. The action of the dipole moment operator $ D= \sum_i\vec d_i\cdot\hat{E} \ket{i}\bra{0}+h.c$ on $\rho_{ss}$ leads, under moderate laser intensity excitation (with a field polarised along the $\hat{E}$ direction), to $D\rho_{ss}= \sum _i\vec d_i\cdot\hat{E}\ket{i}\bra{0}$, which is a superposition of optical coherences $\ket{i}\bra{0}$, i.e, between electronic ground $\ket{0}$ and excited $\ket{i}$ states to be specified in short. Notice that the object $D\rho_{ss}$ displays oscillations with a frequency associated to the energy between ground and excited states. The fast oscillations of the optical coherences (with a period of 2-3 fs) are damped by the  operating dephasing mechanisms (in a 50-100 fs time-scale in physiological environments\cite{EngelNJP,Engel_Nature2006}). These underdamped dynamics can be captured by a master equation  $\partial_t A=-i[{\mathcal H},A]+{\cal L}A={\cal U} A$ with a Hamiltonian part which induces the oscillatory dynamics and a dissipative part in the Lindblad form ${\cal L}A=\sum_k O_kA O_k^+-\frac{1}{2}O_k^+O_kA-\frac{1}{2}A O_k^+O_k$, used for a Markovian (memory-less) relaxation. The coherent evolution of $A=D\rho_{ss}$ is specified  by the Hamiltonian of the interaction between  {\it all} the pigments, arising from the charge redistribution upon excitation of the $Q_y$ transition to the excited state  $\ket{i}$ of each pigment
\begin{eqnarray}
{\mathcal H}&=&\sum_i^N \omega_i\ket{i}\bra{i}+\sum_{i\ne j}^NJ_{ij}(\ket{i}\bra{j}+\ket{j}\bra{i})\nonumber\\
&=&\sum_{\eta=\alpha,\beta}^N\omega_\eta\ket{\eta}\bra{\eta}+\sum_{\substack{\alpha \epsilon RC\\ \beta\epsilon LH}} V_{\alpha,\beta}(\ket{\alpha}\bra{\beta}+\ket{\beta}\bra{\alpha})\nonumber\\
&=& \sum_{\alpha\prime}\omega_\alpha^\prime\Ket{\alpha^\prime}\bra{\alpha\prime}, \label{eq1}
\end{eqnarray} 
that can be expressed in terms of, either, excitonic eigenstates $\ket{\alpha}=\sum_i^{N_{RC}} c_i^\alpha\ket{i}$,  $\ket{\beta}=\sum_i^{N_{LH}} c_i^\beta\ket{i}$ in the absence of the RC-LH coherent interaction $V_{\alpha,\beta}$ and therefore delocalised over either the RC or the LH1, respectively, or, in terms of the full core complex excitonic eigenstates $\ket{\alpha\prime}=\sum_i c_i^{\alpha\prime}\ket{i}$ delocalised over the entire core complex, i.e., over the RC {\it and} the LH1 pigments.  In this article we use primed variables to denote quantities associated to $\ket{\alpha\displaystyle\prime}$ states, greek unprimed for those related  to $\ket{\eta}=\ket{\alpha},\ket{\beta}$ states, and latin letters to denote pigments. The states $\ket{\alpha\prime}$ will be labelled as RC-like or LH1-like states, since they still present a delocalisation which extends mostly over the RC or LH1 pigments,  respectively, given that their mutual coupling $V_{\alpha,\beta}$ is smaller than the energy gap $\omega_\alpha-\omega_\beta$. A realistic model of relaxation processes damping the coherences and broadening the transitions spectra,  the so-called homogeneous broadening, must resort in experimental observations. Three pulse photon echo experiments  \cite{Jimenez_JPCB1997}, underlined pure dephasing (modelled by operators  $O_{\eta\,\epsilon\, \alpha,\nu\,\epsilon\,\alpha}=\sqrt{\gamma_{\eta,\nu}}\ket{\eta}\bra{\nu}$) and intra-ring incoherent transfer dynamics (modelled with $O_{\eta\,\epsilon\, \beta,\nu\,\epsilon\,\beta}=\sqrt{\gamma_{\eta,\nu}}\ket{\eta}\bra{\nu}$ for $\eta \ne \nu$ and rates $\gamma_{\eta,\nu}$   proportional to the spectral density  at the transition frequency $\omega_\eta-\omega_\nu$ and fulfilling detailed balance). Calculations of  absorption spectra were performed by direct Fourier transform  of the evolution super-operator ${\cal F}[e^{{\cal U} t}]=1/(i\omega+{\cal U})$, \cite{Plenio_JCP2013} with averages of stochastic realisations  of  $\omega_i$ and couplings $J_{i,i\pm1}$ in equation (\ref{eq1}), taking into consideration the full set of pigments of the LH1 and $P$. These inhomogeneities, termed static disorder, are complementary to the homogeneous broadening for the full width of optical bands.

A relevant  observation of this article is presented with the red line in Fig.~1B: an important difference  in the absorption spectra between the full core complex ($V_{\alpha,\beta}\ne0$) and the addition of individual RC and LH units ($V_{\alpha,\beta}=0$), is obtained. This difference is  commensurate to the isolated RC spectra, also shown in this Figure, and may arise, both, from the  energy shifts of the frequencies $\omega^{\displaystyle\prime}_\alpha$ with respect to the uncoupled RC and LH1 systems frequencies $\omega_\alpha$, $\omega_\beta$,  and/or, from the increase or decrease in the amplitude of absorption in the full core for specific wavelengths.  The most straightforward comparison to indicate the origin of this difference  is a comparison between the interacting and isolated structures, which in this case, must be addressed by a proper deconvolution of  the full core complex spectrum and a comparison to the isolated  RC and LH1 spectra.  This deconvolution was performed by the partition $D=D_{RC}+D_{LH}$ regarding the dipole operator associated to either RC-like or LH1-like states, respectively,   and the calculation of RC-like and LH1-like DDCF, with a Laplace-Fourier transform of the Markovian master equation evolution kernel\cite{Plenio_JCP2013}  (see SI for further details).

Figures 1C and D show   spectra from isolated RC  and LH1 with a considerably different amplitude than those from the calculated  RC-like  and LH1-like spectra of the interacting core complex, with  an increase of about  60 $\%$  in the area of the P870 band  when the RC is within the core complex. Spectral shifts do not  reflect  the coherent coupling between the RC and LH1, since peak maxima from isolated RC and LH1 spectra  approach to each other in the RC-like and LH1-like spectra,  instead of repelling as is expected from their excitonic interaction $V_{\alpha,\beta}$. The reason is that the change in amplitude is more important for realisations of the core complex where bright RC and LH1 states are closer to resonance, effectively shifting the peak maxima towards the energies lying in between both bands.  
 Equivalent enhancements are obtained in Fig. S4 for different  phenomenological  dephasing models, which stand for a site dephasing (full homogeneous width given by  operators $O_i=\sqrt{\gamma_i} \ket{i}\bra{i}$) or collective dephasing scenarios (constructed by operators $O_{\alpha\prime}=\sqrt{\gamma_\alpha^{\displaystyle\prime}}\ket{\alpha^{\displaystyle\prime}}\bra{\alpha^{\displaystyle\prime}}$ for the dephasing contribution). The equivalence of absorption spectra resulting from different dephasing models illustrates the robustness of the increase of the P870 band and the possibility  to trace back such an increase  by any of the models studied. In particular, in the collective dephasing model, DDCF$(t)=\sum_{\alpha} | D_{\alpha}^{\displaystyle\prime}|^2 e^{(i\omega_\alpha^\prime-\gamma_{\alpha}^\prime})t$, which results in an absorption spectrum $A(\omega)=\sum_{\alpha\displaystyle\prime} \gamma|D_\alpha^{\displaystyle\prime}|^2 /[\gamma^2+(\omega-\omega_\alpha^{\displaystyle\prime})]^2$ for a single realisation within the inhomogeneous ensemble. Absorption spectra  amplitudes are proportional to the  dipole strengths   of the RC-like or LH1-like states $| D_{\alpha}^{\displaystyle\prime}|^2=|\sum_i c_i^{\alpha\displaystyle\prime} \vec d_i\cdot \hat{E}|^2$, hence, the increase of the absorption cross section is closely related to the increase the the RC-like states dipole strength $| D_{\alpha}^{\displaystyle\prime}|^2$. 

In order to foresee the elements that support this robust increase associated to  $| D_{\alpha}^{\displaystyle\prime}|^2$, for the moment, let us consider a single bright state of the P870 band and a single bright state of the B880 band, which couple according to the  Hamiltonian 
\begin{eqnarray}
H&=&\Delta E\ket{P870}\bra{P870}+V(\ket{P870}\ket{B880}+\nonumber\\
&&\ket{B880}\bra{P870}) \label{redH} 
\end{eqnarray}
leading to the RC-like dipole strength $| D_{P870}^{\displaystyle\prime}|^2$,
\begin{eqnarray}
 | D_{P870}^{\displaystyle\prime}|^2 = |\cos\theta D_{P870}+\sin\theta D_{B880}|^2\label{dip_red}
\end{eqnarray}
where the mixing angle $\theta=\frac{1}{2} \arctan (2V/\Delta E)$ (here $\Delta E=\omega_{P870}-\omega_{B880}$). For moderate couplings $V/\Delta E\ll 1$ the  redistribution of absorption amplitude   $ | D_{P870}^{\displaystyle\prime}|^2- |D_{P870}|^2\approx 2 \mbox{ Re}\{D_{P870}^* D_{B880}V/\Delta E\}$  depends on  an interference term with a more favourable scaling $\propto V/\Delta E$, than that associated to the energy shifts $\Delta E^{\displaystyle\prime}-\Delta E\approx 2 V^2/\Delta E $. The increase  $ | D_{P870}^{\displaystyle\prime}|^2- |D_{P870}|>0$ of the P870 absorption occurs in the core complex as a consequence of a higher energy of the RC P870 with respect to the LH1 B880, $\Delta E>0$, and the inequality Re$\{( D_{P870}^*  D_{B880}) V\}>0$ that results from the almost tangential arrangement of LH1 BChl transition dipoles  \cite{karrasch95,Cogdell_Science_2003} (see SI for further details). Moreover, thanks to this interference term, the RC transition suffers a change proportional to the harvesting transition dipole and benefits thereby from mechanisms that amplify this dipole strength, e.g. excitonic delocalisation over several harvesting pigments or intensity borrowing from other transitions of individual pigments to the $Q_y$ band, which are both  patent in the {\it R. rubrum} LH1 complex \cite{vanGrondelle_1997JPCB,Ghosh_1988BioC}.

We have been able to numerically ascertain that  the absorption redistribution is additive  (whenever $\theta\ll 1$, see SI), i.e., the amplification of the transition dipole of the RC  is  very similar when arising from a set of LH states or from the addition of the contribution of each transition of this set. Altogether, excitonic delocalisation, $Q_y$  intensity borrowing, and additivity of the important set of bright states in the LH1 under static disorder, support  the important increase of the P870 absorption strength observed in Fig.1C.

\begin{figure}
\centering
\includegraphics[width=.51\columnwidth]{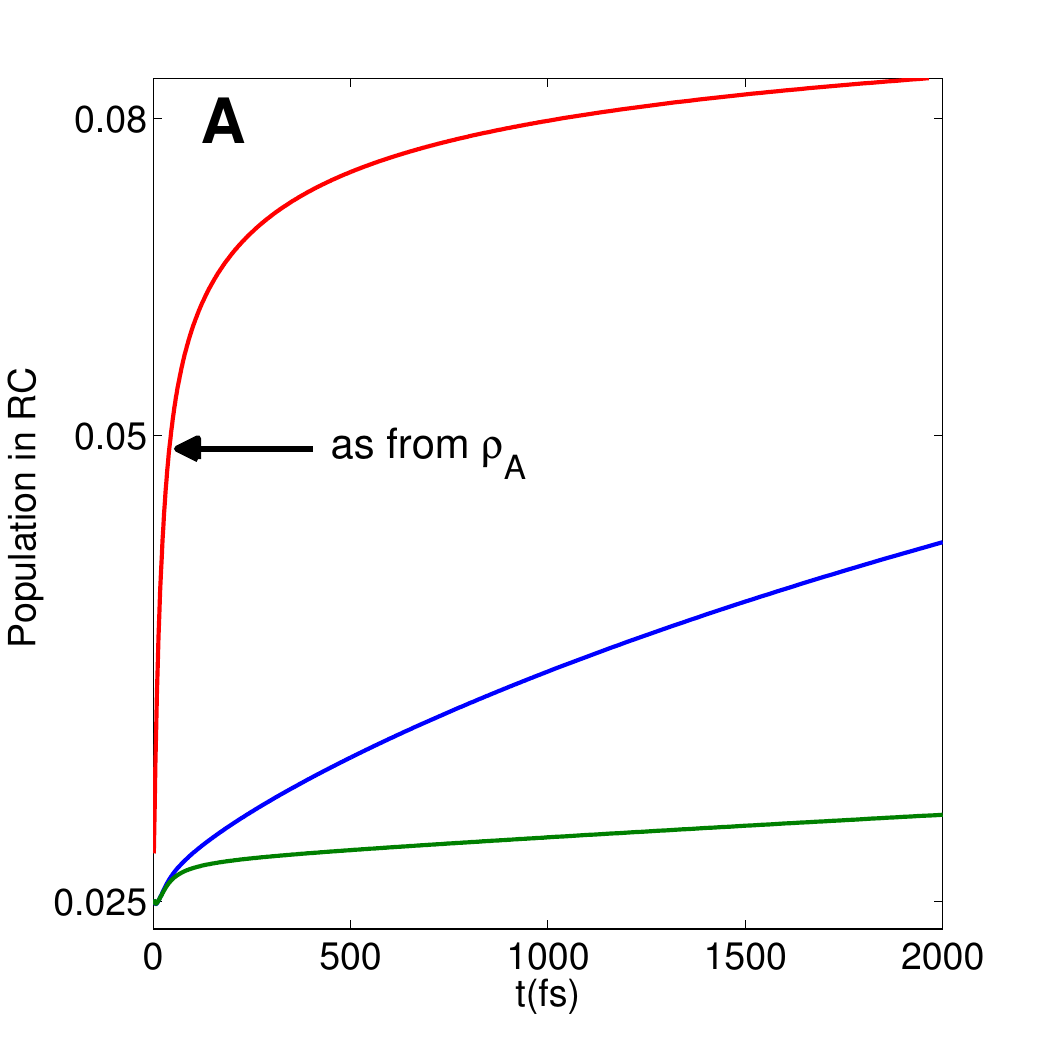}\hspace{-.3 cm}
\includegraphics[width=.45\columnwidth]{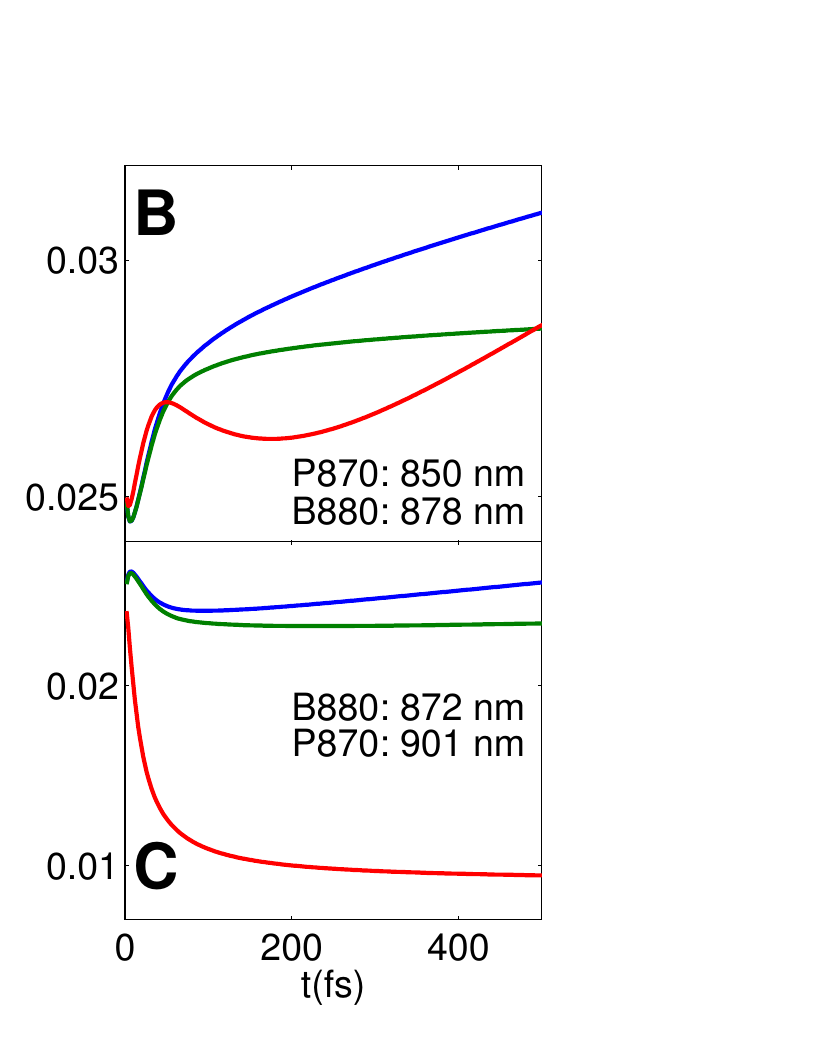}
\caption{ {\bf Population  dynamics in the core complex under incoherent light illumination.} {\bf A} Ensemble population of the $P$ pigments ($10^3$ realisations of equation (1)). The arrow shows the theoretical estimate for the maximum coherent population redirection to the RC according to $\rho_A\propto \sum_{\alpha^{\displaystyle\prime}}| D_{\alpha}^{\displaystyle\prime}|^2\ket{\alpha^{\displaystyle\prime}}\bra{\alpha^{\displaystyle\prime}}$. {\bf B} and {\bf C} Individual realisations of the ensemble where the P870 lies with an energy above  and  below the B880 band, respectively.  The identification of B880 or P870 is made based on the brightest state $\ket{\alpha}$ associated to either RC or LH1, from  diagonalisation of $H$ with $V=0$, and respective energies that correspond to the wavelength in the inset of each figure. Green, blue and red correspond to site dephasing, partial excitons dephasing and collective dephasing models, as explained in the main text. }\label{fig2}
\end{figure}

The main dynamic consequence  of  this absorption redistribution is the phenomenon we call  population redirection,  which stands for a redistribution of excitations that reflects the coherent interaction between pigments, albeit the incoherent nature of  illumination (see SI for further details of the model). This redirection is illustrated in Fig.~2A.  The incoherent illumination excites the  $P$  pigments  according to their individual  dipole strength $\propto\sum_{i\epsilon RC} |\vec d_i\cdot \hat{E}|^2$ at very early times $t\approx 0$ fs. Then, follows population dynamics {\it towards} the RC occurring on a time-scale shorter than the inverse of the effective dephasing rate which is, thereby, partially coherent. In the absence of any relaxation, this redirection would not be possible since populations would display coherent reversible oscillations between RC and LH pigments,  with a  period  given  by $t^*=2\pi/\sqrt{\Delta E^2+4V^2}$  in our reduced two level model. Thereby, an initial pull driven by coherent dynamics becomes directed, hence irreversible, due to the  establishment of incoherent dynamics, manifest by the change of slope common to  all curves of Fig.2 after 120-150 fs.
Since $\Delta E\gg V$ in the core complex, this population increase is fast thanks to the energy mismatch, $\Delta E$, but nevertheless originated by $V$.
This transient can be visualised as the result of a very early absorption on pigment excited states $\ket{i}$, after which, the excitonic interactions and the dephasing dynamics mold the states that will dominate the classical dynamics, i.e. the incoherent transfer process.
Under this representation, it is legitimate  to describe the density operator  that emerges by the coherent-incoherent interplay in the collective dephasing model, by  $\rho_A=\sum_{\alpha^{\displaystyle\prime}} | D_{\alpha}^{\displaystyle\prime}|^2\ket{\alpha^{\displaystyle\prime}}\bra{\alpha^{\displaystyle\prime}}/I$,  built upon the same weights that describe the contributions to the DDCF. The validity of this ansatz is supported by Fig.2A, which shows that the resulting $P$ population from $\rho_A$ (arrow), is in agreement with the predicted population for the collective dephasing model when  the coherent-to-incoherent  transition starts to occur.  Since each dephasing model will mold a particular einselection that presents different characteristics \cite{Zureck_RMP2003}, unsurprisingly, the site and realistic dephasing models present a different amount of population redirection (15\%$P$ population increase) than the collective dephasing (80\%$P$ population increase)  after a time  $\simeq 100$ for the ensemble average population presented in Fig.2A. Notice that  the population redirection  occurs {\it towards} the RC  when $\Delta E>0$ or  {\it away} from the RC when $\Delta E<0$, as can be observed in Figs.~2B and C for specific realisations of the ensemble that present such energy landscape. 
Therefore, the rather robust absorption redistribution towards the RC band in the core complex shown in Figs.~1B and S4, represents a clear  signature of a more delicate, but nonetheless persistent, coherent population redirection towards the RC pigments under natural illumination.

\paragraph*{Predicted absorption redistribution in general photosynthetic structures: uncovering principles  from optimisation of artificial light harvesters}

The possibility to  quantify  the absorption redistribution based on the dipole strengths $| D_\alpha^{\displaystyle\prime}|^2$ as weights, allows us to predict an analogous  absorption spectra redistribution  in photosystem (PS) 1 and PS2 monomer  of higher plants, with a 37$\%$ and 50\% increase of the $P$ band, respectively,  due to excitonic delocalisation over antenna pigments. The natural dimeric structure of PS2 presents, however, almost no enhancement, and hence no potential redistribution to the RC pigments   (see SI for additional details). The prediction of absorption redistribution, and hence, the possibility of an important population redirection -subject  to  the dephasing phenomenology- which is conserved across some species, draws our attention to further unveil the principles that stand behind this coherent effect under physiological conditions, and understand how it can be improved for guiding the design of artificial light harvesters. Thus, we have developed an optimisation procedure of the population of a single target pigment according to $\rho_A$ based on stochastic variations of positions and orientations of a set of $N$ identical harvesters (further details in SI). Figure 3A shows that the dipole strengths $|{D_{LH}}|^2$ associated to the $N$ harvesters of the optimal configurations,  is concentrated almost entirely in a single transition $\ket{\eta}$ such that $D_{LH}=\sum_{i}^N |\vec d_i\cdot\hat{E}|^2 \simeq | D_{\eta}|^2$, with the remainder of states  therefore dark, $| D_{\nu\ne\eta}|^2\approx 0$.  This finding permits to describe the optimal configurations by a simple two level system as described by a Hamiltonian equation (\ref{redH}), and which under a collective dephasing, will develop a population  in the target state  $\ket{T}$,  
\begin{eqnarray}
P_T &=& \frac{1}{4I} \bigg\{ |{D}_T|^2(3+\cos4\theta)+2\Re\left\{{D}_T^*{D}_{LH}\right\}\sin4\theta\nonumber\\
&&+|{D}_{LH}|^2(1-\cos4\theta) \bigg\} \label{eqChris}
\end{eqnarray}
expressed in terms of the target and light-harvesting dipoles ${D}_T$ and ${D}_{LH}$. This population,  for $\tan(4\theta^{max})=2\Re \{ D_T^* D_{LH}\}/ (| D_T|^2-| D_{LH}|^2)$, acquires a maximum value 
\begin{eqnarray}
P_T^{max}=\frac{2| D_T|^2+| D_{LH}|^2}{2I}\label{eqChris2}
\end{eqnarray} 
where $I=| D_T|^2+| D_{LH}|^2$ is, as before, proportional to the total absorption  cross section of the device (details of the calculation are presented in the SI). Whenever $N\gg1$, $| D_{LH}|\gg | D_{T}|$  and  equation (\ref{eqChris2}) predicts that  {\it half} of the absorbed photons will be redirected to the target pigment, corroborated by our optimal configurations already for $N=7$, in  the inset of Fig.~3A.
The scope of these results is easier to grasp with an example: an optimal device with 100 harvesting units will be able to empower a 50-fold increase of the population at a {\it single} target unit.

\begin{figure}
\centering
\begin{minipage}{0.55\columnwidth}
\includegraphics[width=1\columnwidth]{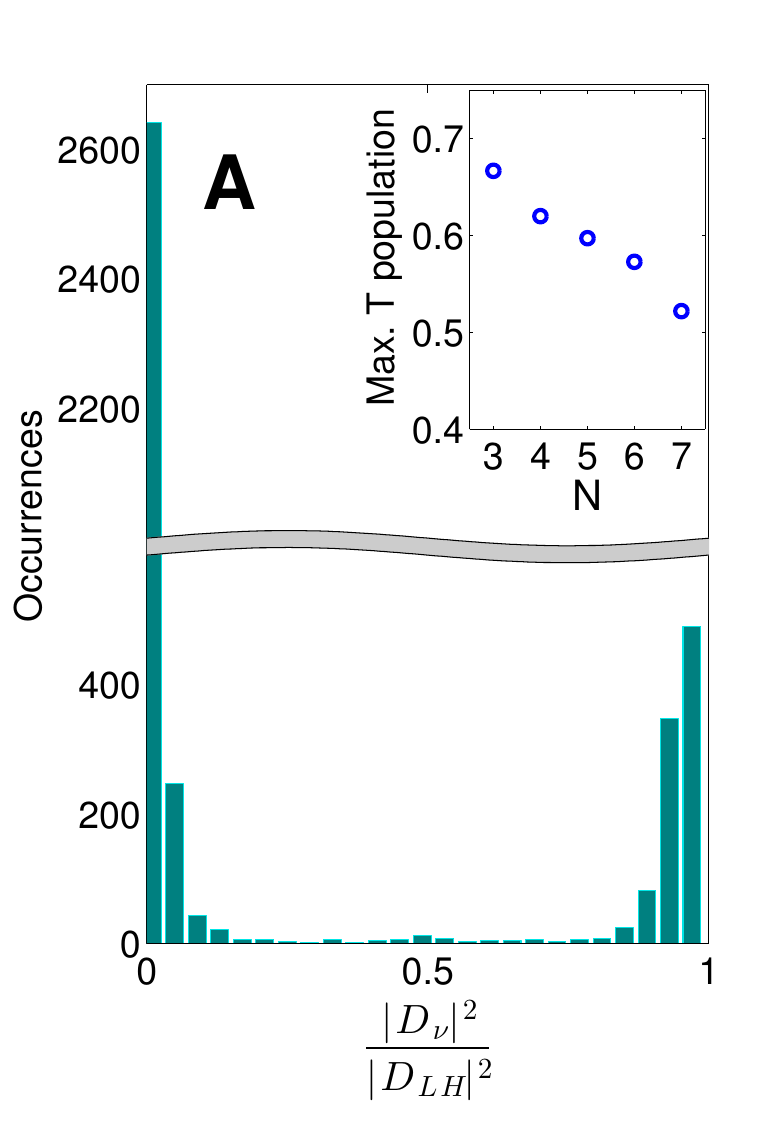}\\
\end{minipage}
\begin{minipage}{0.4\columnwidth}
\includegraphics[width=1\columnwidth]{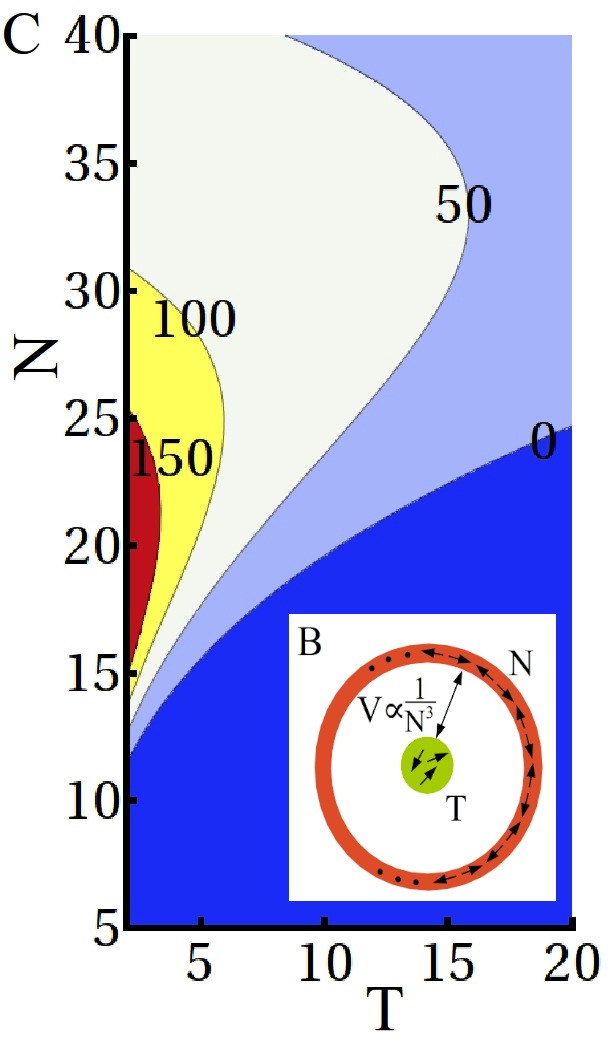}
\end{minipage}

\caption{ {\bf Design principles for guided coherent redirection to target pigments.} {\bf A} Histogram of dipole strengths from the states of 800  configurations that optimise the population of a single target pigment according to $\rho_A$, due to the coherent interaction with five harvesting pigments. Each of  these optimal  configurations was obtained after  $10 ^5$ trials of the position of the harvesting pigments and the orientation of their transition dipoles. The inset shows the absolute maximum population of the target pigment from these configurations as a function of the size of the harvesting structure $N$. Using a simplified model for a general circular aggregate {\bf B}, the percentage increase of the initial excitation of the $P$-pigments is calculated in {\bf C} as a function of the number $T$ of  target units and number $N$ of LH pigments. The excitation of target units is greater for $T\ll N$ and is maximised for $N \sim 20$. Peak absorption wavelengths of target and harvesting transitions at 880 and 870 nm, respectively. }
\end{figure}

However, an optimal redirecting device with $N\gg 1$ placed in a geometry that develops  a single bright transition, naturally increases the distance of the LH to the target pigments, reducing the coherent coupling $V$. To meet this practical constraint we study a geometrical design in Fig.~3B that locates target units encircled by harvesters, these latter separated by a constant arc length. A balance between  the  transition dipole strength $|D_{LH}|^2\propto N$, and the magnitude of the target-harvesters coupling  $V\propto 1/N^3$,  results in an optimal population redirection for a finite $N$ (150\% enhancement of $T$ population for $N=21$ in Fig.~3C).

The design of artificial devices can integrate both coherent redirection and incoherent relaxation to grant excitation of the target pigment, that can be achieved with an energy landscape where the excited state of the target pigment has a lower energy than the harvesting transitions. Large harvesting structures that encircle a target pigment  $| D_{LH}|^2\gg | D_{T}|^2$ will present a  small coupling $V$, but as in the core complex, they must ensure an important energy mismatch $|\Delta E|$ to drive the population redirection before excitonic decoherence sets. This results in a mixing angle that obeys both $\tan(4\theta^{max})\approx 4 V/\Delta E$ and  $\tan(4\theta^{max})\approx - 2| D_T|/| D_{LH}|$, thereby with an optimal energy mismatch $\Delta E\simeq-2V| D_{LH}|/| D_{T}|$, When   $V\gtrsim 0$ the  target transition presents a  lower  energy than the main harvesting transition with dynamics that profit from both coherent population redirection and  the usual incoherent relaxation funnelling --via the F\"orster energy transfer mechanism-- in order to populate the target pigment excited state.

In conclusion, theoretical considerations have led us to identify a new design principle which exploits an interplay between light absorption and excitonic delocalisation, in order to produce a coherent dynamics which redirects excitation  to the RC in several photosynthetic systems. 
Our detailed theoretical analysis elucidates how the optical response  and size of the harvesting units, energy mismatch between these and target pigments, as well as system inhomogeneities, are important figures of merit to set clear guidelines for the design of light-harvesting devices that efficiently direct photon absorption by means of coherent excitonic dynamics.

We acknowledge discussions with Fedor Jelezko, James Lim and Shai Machnes. This work was supported by an Alexander von Humboldt Professorship, the ERC Synergy grant BioQ, the EU STREP PAPETS and the DFG SFB TRR 21. Additional support was provided by the National Science Foundation through PFC@JQI and the National Science Foundation Graduate Research Fellowship Program under DGE-1322106.

\clearpage
\appendix
{\centering{\bf Supplementary Information}}
\section{LH1 structure, Hamiltonian and analytic results}\label{sec:H}

\begin{figure}[ht]
\includegraphics[width=0.5\columnwidth]{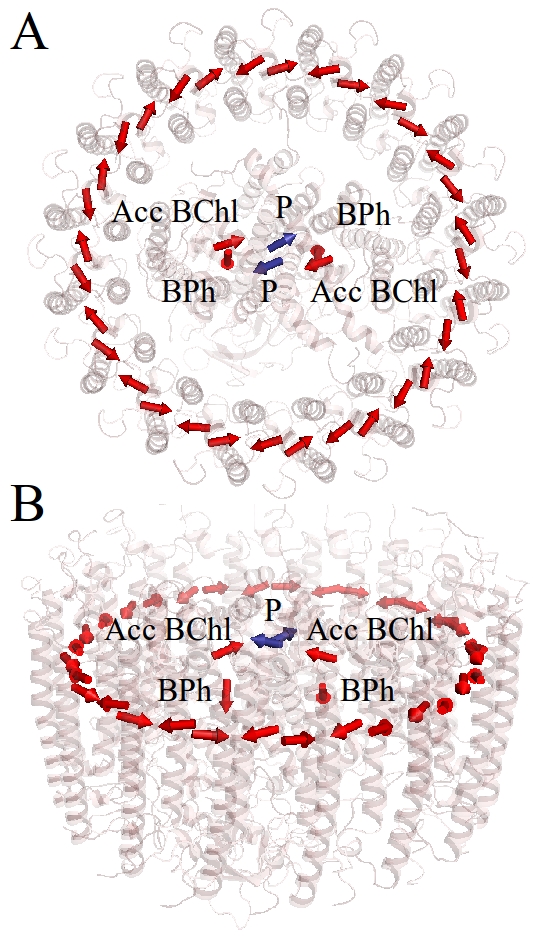}
\caption{({\bf A}), ({\bf B}) The position of the BChl and bacteriopheophytin pigments in the atomic model of the {\em R. rubrum} LH1-RC complex used here. The orientations of the $Q_y$ transition moments of the pigments are shown with arrows. The special pair ($P$), accessory BChls ($B_L$ and $B_M$) and bacteriopheophytins ($H_L$ and $H_M$) of the RC are also indicated.  The $\alpha$ and $\beta$ protein subunits are rendered translucent.} \label{fig:PSU_schematic}
\end{figure}

In this study, we have employed the assumption of $C_{16}$ symmetry for the structure of the LH1, based upon  various experimental observations \cite{Gerken_JPC2003,Ghosh_Biochem_2003} and recent published x-ray structure data \cite{Niway_2014Nat}. Its pigments are identified by indices $(m,\zeta)$, where $m \in \{-N/4+1,\ldots,N/4 \}$ labels the (dimerized) unit cell ($N=32$), and $\zeta \in \{1,2 \}$ labels the pigment within the $\alpha\beta$ heterodimer. Low energy spectral features can be characterised using the $Q_y$ optically-induced transition dipole moments of the BChl, which lie along a line joining the nitrogens of unreduced pyrrole rings I and III \cite{Weiss72}. A general parameterisation of the LH1 BChl positions of {\it R. rubrum} was obtained from the agreement with a pdb file obtained by R. Ghosh\cite{Autenrieth_2002} in a collaboration with the Beckmann Institute, University of Illinois, Urbana-Champaign. This parametrisation of the LH1 BChl position $\vec{r}_{m,\zeta}$, and the corresponding $Q_y$ transition dipole moments, $\vec{d}_{m,\zeta}$, is: 
\begin{align}
\vec{d}_{m,\zeta} &= d 
	\begin{pmatrix}
		- (-1)^\zeta \sin \left(\frac{4\pi}{N}m + \frac{\gamma}{2}(-1)^\zeta + \Delta\gamma_\zeta\right) \cos \phi_\zeta \\
		 (-1)^\zeta \cos\left(\frac{4\pi}{N}m + \frac{\gamma}{2}(-1)^\zeta + \Delta\gamma_\zeta\right) \cos \phi_\zeta \\
		\sin \phi_\zeta \\
	\end{pmatrix} \nonumber\\
\vec{r}_{m,\zeta} &= \begin{pmatrix}
		r_\zeta \cos\left(\frac{4\pi}{N}m + \frac{\gamma}{2}(-1)^\zeta\right)  \\
		r_\zeta \sin\left(\frac{4\pi}{N}m + \frac{\gamma}{2}(-1)^\zeta\right)  \\
		(-1)^\zeta z/2 \\
	\end{pmatrix}.\label{paramet}
\end{align}
An explanation of structural parameters and the values used in the full numerical study are summarised in Table S1.

A correspondence between microscopic structure and optical properties is possible by a theoretical analysis using a Hamiltonian coupling between $Q_y$ transition dipoles. We consider the Hamiltonian:
\begin{eqnarray}
    \mathscr{H} &=&\sum_{m,\zeta} \epsilon_{m,\zeta} \ket{m,\zeta}\bra{m,\zeta}+\nonumber\\
&&\sum_{(m,\zeta)\ne(m',\zeta')} J_{m,\zeta}^{m',\zeta'} \ket{m,\zeta}\bra{m',\zeta'}+ H.c. \label{eq:H}
\end{eqnarray}
where $\epsilon_{m,\zeta}$ is the site energy of BChl $(m,\zeta)$ and $J_{m,\zeta}^{m',\zeta'}$ is the interaction energy between pigments $(m,\zeta)$ and $(m',\zeta')$. When the details of the charge distribution of excited states are of no importance, $J_{m,\zeta}^{m',\zeta'}$ can be calculated using the point-dipole approximation, whereas 
the sub-nanometer distance between neighbouring chromophores implies that the nearest-neighbour couplings depend on the geometry of the electronic wave-function of each chromophore. 
 
The Hamiltonian of equation (\ref{eq:H}) is best examined in the Fourier basis
\begin{align}
    \ket{k,\zeta} = \frac{1}{\sqrt{N/2}} \sum_m  e^{i \frac{4\pi}{N} k m} \ket{m,\zeta}
\end{align}
where $k \in \{-N/4+1,\ldots,N/4 \}$. In this basis, the states decouple for different values of $k$
\begin{align}\begin{split}
    \bra{k,\zeta} \mathscr{H} \ket{k',\zeta'} &= \mathscr{L}_{\zeta,\zeta'}(k) \delta_{k,k'} \\
    \mathscr{L}_{\zeta,\zeta'}(k) &= \sum_m  e^{i \frac{4\pi}{N} k m} \bra{0,\zeta}\mathscr{H}\ket{m,\zeta'}.
\end{split}\end{align}
The diagonalization of the $2 \times 2$ Hamiltonians restricted to each $k$-subspace, $\mathscr{L}(k)$, is 
accomplished with the unitary:
\begin{align}
    U(k) =
    \begin{pmatrix}
	   	   e^{i \Phi(k)/2}\cos{\Theta(k)} & e^{i \Phi(k)/2}\sin{\Theta(k)} \\
		  -e^{-i \Phi(k)/2}\sin{\Theta(k)} & e^{-i \Phi(k)/2}\cos{\Theta(k)} \\
    \end{pmatrix}.
\end{align}
where $\Phi(k) = \arg(\mathscr{L}_{01}(k))$ and $\Theta(k) = \nicefrac{1}{2} \arctan{ \frac{2|\mathscr{L}_{01}(k)|}{\mathscr{L}_{00}(k) - \mathscr{L}_{11}(k)} }$ reflect the ``amount of dimerisation'': $\Phi(k)$ captures the difference between intra- and inter-dimer coupling, and $\Theta(k)$ captures the site energy differences.

Considering degenerate pigments $\epsilon_{m,\zeta}=\epsilon$ and taking into account only nearest neighbour coupling, the matrix $\mathscr{L}$ takes on a particularly
simple form:
\begin{align}
\mathscr{L}(k) &=
\begin{pmatrix}
		\epsilon & Q_1+Q_2 e^{i\frac{4\pi}{N} k} \\
		Q_1+Q_2 e^{-i\frac{4\pi}{N} k} & \epsilon \\
\end{pmatrix}
\end{align}
where $\epsilon$, $Q_1$, $Q_2$  are  site energies, and intra-dimer and  inter-dimer couplings, respectively. Accordingly:
\begin{align}\begin{split}
\Theta(k) &= \frac{\pi}{4} \\
\Phi(k) &= \arctan \frac{Q_2 \sin \frac{4\pi}{N} k}{Q_1+Q_2 \cos\frac{4\pi}{N} k}.\label{eq:davphi}
\end{split}\end{align}
The Hamiltonian is diagonalized by the exciton states $\ket{k,\nu}=\sum_{m,\zeta} c_{m,\zeta}^{k,\nu} \ket{m,\zeta}$ with energies $\hbar\omega(k,\nu)$ 
\begin{align}
\ket{k,\nu} &= \sum_{m,\zeta} c_{m,\zeta}^{k,\nu} \ket{m,\zeta}
             = \frac{1}{\sqrt{N/2}}\sum_{m,\zeta} e^{i \frac{4\pi}{N} k m} \, U(k)_{\zeta,\nu} \ket{m,\zeta}\label{states}\\
\omega(k,\nu) &= \epsilon+(-1)^\nu \left( Q_1 + Q_2\cos \frac{4\pi}{N} k  \right), \label{eq:davenergy}
\end{align}
The circular symmetry is manifest in the two-fold degeneracy $\omega(k,\nu)=\omega(-k,\nu)$.

Without static noise, the $P870$ band would stand form a single state $\ket{P870}=\nicefrac{1}{\sqrt{2}} \, \left( \ket{P1}-\ket{P2} \right)$ since the dipole moments of the $P$-pigments with degenerate energies are nearly anti-parallel. This bright special pair state is modelled by:
\begin{align}
\vec{r}_{P870} =
	\begin{pmatrix}
		0 \\
		0 \\
	\end{pmatrix} \qquad
\vec{d}_{P} =|\vec{d}_P|
	\begin{pmatrix}
		\cos\beta \\
		\sin\beta \\
	\end{pmatrix} \qquad
\end{align}
where $T=2$ are the number of pigments associated to the P870 transition, while $|\vec{d}_P|^2$ are their individual dipole strengths,  and $\beta$ parametrize the transition dipole direction of the $\ket{P870}$ state. 

The coupling between LH1 excitons and $P$-pigments, is then, after some algebra using the point-dipole approximation between RC and LH1 pigments for their excitonic interaction $J$:
\begin{eqnarray}
V_{k,1} &= \bra{P870}J\ket{k,1}=\delta_{k,\pm1}\frac{\sqrt{N T}}{2} \frac{d_{LH}d_{P}}{R^3} \cos\left(\frac{\Phi(1)+\gamma}{2}\right) \nonumber\\
    &  \qquad \times (\mp i)e^{\pm i(\beta-\Delta\gamma)}\left(1 \pm 3ie^{\pm i\Delta\gamma}\sin\Delta\gamma \right) \label{eq:vk}\\
V_{k,2} &= \bra{P870}J\ket{k,2}=\delta_{k,\pm1}\frac{\sqrt{N T}}{2} \frac{d_{LH}d_P}{R^3} \sin\left(\frac{\Phi(1)+\gamma}{2}\right) \nonumber\\
    &  \qquad \times (\mp i)e^{\pm i(\beta-\Delta\gamma)}\left(1 \pm 3ie^{\pm i\Delta\gamma}\sin\Delta\gamma \right), 
\end{eqnarray}
where $R$ is the radius of the LH1 ring. These expressions illustrate a selection rule, $V_{k,\nu} \propto \delta_{k,\pm1}$, for the coupling between LH1 excitons and the $P$-pigments.  Increasing dimerisation, $Q_1>Q_2$, leads to $\Phi(1)+\gamma\ll1$ and, consequently, a stronger coupling to the P870 state from the lower energy band $\nu=1$.

It is interesting to notice  that increased dimerization enhances the coupling strength  between the RC and the lower energy manifold $\nu=1$ in the LH1, and consequently,  the coupling between $\ket{P870}$ and the $\ket{\pm1,1}$ state. Dimerisation also places all $P$ dipole strength on the $\ket{P870}$ state, which also represents the lowest energy state of this coupled dimer. Enhancements of the dipole strength of the low energy states ensure thermalization on highly interacting states, boosting incoherent F\"orster rates.

For a Hamiltonian description of the symmetric situation, we notice that $\ket{P}$ only  couples to $\ket{k=\pm1,1}$, thereby, we need only three-levels in order to characterise RC-LH1 interactions, described by the Hamiltonian 

\begin{align}
\mathscr{H}&=
	\begin{pmatrix}
		\Delta E &  V_1  & V_{-1} \\
		V_1^*      &  0  & 0 \\
		V_{-1}^*        &  0  & 0 \\
	\end{pmatrix}.
\end{align}
where $V_1=V_{1,1}$ of equation (\ref{eq:vk}). This three-level system can be reduced to an effective two-level system by choosing a suitable basis for the degenerate $k=\pm1$ subspace
\begin{align}
    \ket{B880} &= \frac{1}{\sqrt{2}} \left( e^{-i\psi}\ket{k=1} + e^{i\psi} \ket{k=-1} \right), \\
    \ket{\emptyset} &= \frac{1}{\sqrt{2}} \left( e^{-i\psi}\ket{k=1} - e^{i\psi} \ket{k=-1} \right)
\end{align}
where $\psi=\arg(V_1)$. The 
Hamiltonian becomes
\begin{eqnarray}
\mathscr{H}&=
	\begin{pmatrix}
		\Delta E  &  V  & 0 \\
		V         &  0  & 0 \\
		0         &  0  & 0 \\
	\end{pmatrix},\label{eq:H_TLS}
\end{eqnarray}

which is identical to the two-level Hamiltonian of equation (2) in the main text. Diagonalization of equation (\ref{eq:H_TLS}) leads to delocalized eigenstates
\begin{eqnarray}
\ket{P870^{\displaystyle \prime}}  &= \cos\theta \ket{P870} + \sin\theta \ket{B880}, \label{eq:app_Pwave}\\
\ket{B880^{\displaystyle \prime}}  &=-\sin\theta \ket{P870} + \cos\theta \ket{B880}, \label{eq:app_Bwave}\\
\theta &= \frac{1}{2} \arctan\left( \frac{2V}{\Delta E} \right).
\end{eqnarray}

and to the modified dipole strength of equation (3) of the main text. Given that  $\Delta E \equiv 130$ cm$^{-1}$ and $V=\sqrt{2}|V_1| \approx 13$ cm$^{-1} \ll \Delta E$, at first order in  $\nicefrac{V}{\Delta E}\ll1$    we obtain
\begin{eqnarray}
 |{D}^{\displaystyle \prime}_{P870}|^2              &\approx|{D}_{P870}|^2 + 2\Re\left\{{D}_{P870}^*{D}_{B880}\right\} \left(\frac{V}{\Delta E} \right)  \label{eq:app_eq2} 
\end{eqnarray}
where $D_{P870} =  \vec d_P \cdot\hat{E}$ stands for the component of $\vec d_P$ along the exciting polarisation axis.
In accordance with the dipole sum rule, the B880 band strength
\begin{eqnarray}
|{D}^{\displaystyle \prime}_{B880}|^2 &=|{D}_{B880}|^2 \cos^2\theta - 2\Re\left\{{D}_{P870}^*{D}_{B880}\right\} \cos\theta\sin\theta \nonumber\\
& \qquad + |{D}_{P870}|^2 \sin^2\theta
\end{eqnarray}
changes by an amount such that the total absorption intensity $|{D}^{\displaystyle \prime}_{B880}|^2+|{D}^{\displaystyle \prime}_{P870}|^2=|{D}_{B880}|^2+|{D}_{P870}|^2$ remains constant.

To proceed, it is necessary  to calculate the transition dipoles $D_{B880}$, which follow from $D_{k,\nu} = \sum_{m,\zeta} c_{m,\zeta}^{k,\nu} \vec{d}_{m,\zeta}\cdot{\hat E}$ 

\begin{eqnarray}
    D_{k,1} &= \mp i\delta_{k,\pm1}d_{LH} \sqrt{N} \cos\left( \frac{\Phi(1)+\gamma}{2} \right) \frac{e^{\pm i (\Delta\gamma+\delta)}}{2}
    \label{eq:dip1}\\
    D_{k,2} &= \delta_{k,\pm1} d_{LH} \sqrt{N} \sin\left( \frac{\Phi(1)+\gamma}{2} \right) \frac{e^{\pm i (\Delta\gamma-\delta)}}{2}
      \label{eq:dip2}
\end{eqnarray}
where $\gamma$ is the intra-dimer angle, $\Delta \gamma$ is the tilt of the dipoles away from the tangent, as described in Table S1, and $\delta$ is the polarisation $\hat {E}$ angle of the incident electric field.  The concentration of dipole strength in the lower bands is a function of $\Phi(1)+\gamma$, which reflects the amount of dimerisation. According to our model of the LH1, more than $\gtrapprox 95\%$ of the total absorbing strength is in the low-lying $\ket{\nu=1}$ manifold, which is responsible for the strong response of the LH1 at $\lambda= 880$ nm$\approx 2\pi/\omega(\pm1,1) $.   The fact that only the low energy lying states couple considerably to the P870 transition can be viewed as a consequence of the point dipole approximation, since it states that the only states of the ring that couple to the RC are those having a non-vanishing transition dipole.

These transition dipole components can be inserted in equation (\ref{eq:app_eq2}), taking care on following the transformation of equations (\ref{eq:app_Pwave}) and (\ref{eq:app_Bwave}). An additional average over all orientations of the electric field polarisation $\hat{E}$ result in a redistribution of transition dipole strength for the P870 state described by  
\begin{eqnarray}
|{D}^{\displaystyle \prime}_{P870}|^2  \approx|{D}_{P870}|^2 &\left(1 + \frac{N}{\sqrt{2}} \frac{d_{LH}^2}{R^3\Delta E} \cos^2\left( \frac{\Phi(1)+\gamma}{2}\right)(1-3\sin^2\Delta\gamma) \right). \label{eq:app_e2_ext}
\end{eqnarray}

Equation \ref{eq:app_e2_ext} explicitly shows that, as long as $\Delta E>0$,  
the bright $P870$ state increases its absorption intensity for small $\Delta\gamma$ (transition dipoles nearly tangent 
to the ring) and $\Phi(1)+\gamma$ (dimerized unit cell). The uphill energy landscape, $\Delta E>0$, which sets transfer-to-trap as the bottleneck in the core complex, increases direct absorption in the $P870$ band.

\section{Approximate additivity of absorption redistribution}
One  of the interesting facts of the absorption redistribution is that a system consisting of a single target excited state  which couples to  many excitonic harvesting transitions, will develop a transition dipole which is is given approximately  by the addition of all contributions from each of the individual harvesting transitions.

To illustrate this fact, we consider the pairwise  Hamiltonian $H_i=-\Delta E  \sum_i\ket{LH_i}\bra{LH_i}+v_i (\ket{LH_i}\bra{T}+\ket{T}\bra{LH_i})$ regarding a given harvesting transition $\ket{LH_i}$ having an energy $-\Delta E$ with respect to the  target transition $\ket{T}$.  Diagonalization of this Hamiltonian $H_i$ results in eigenstates with transition dipoles of target-like states (states which are mainly delocalised over the target transition) $D_T$, which differ from the isolated target transition dipole $d_T$, by the an absorption strength $\Delta D_i=(D_{T,i}^2-d_T^2)/d_T^2 $, normalised by the the absorption cross section of the isolated target $d_T^2$. For a system of $N$ pairs of harvesting and target transitions, the addition $\sum \Delta D_i$ will show how much the absorption cross section of  the target pigment transitions, is increased.
 
\begin{figure}
\includegraphics[width=.38\columnwidth]{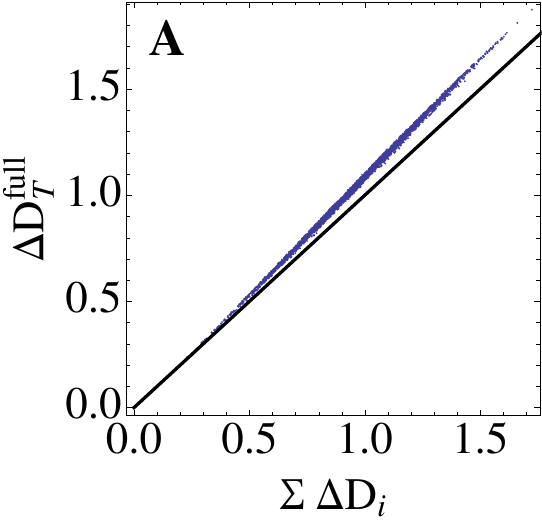}
\includegraphics[width=.41\columnwidth]{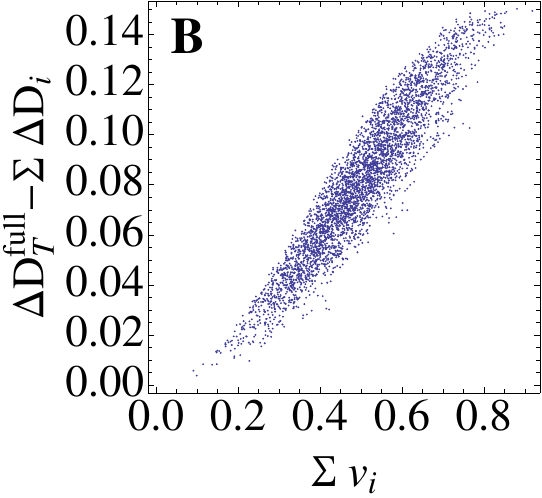}
\caption{Additivity of absorption redistribution for moderate excitonic couplings. {\bf A} shows the scatter of events of the full redistribution of absorption $\Delta D_T^{full}$ and the addition of all individual harvesting-target pairs $\sum_i\Delta D_i$ from a system of $N=6$ harvesting transitions. The events represent the output  from realisations of values $v_i$  uniformly distributed within [0,$\Delta E/5$], hence $\langle v_i\rangle= \Delta E/10$ as is the case for the core complex of {\it R. rubrum}. The line guides the eye for the equality  $\sum_i\Delta D_i=\Delta D_T^{full}$. In {\bf B} the scatter points show the positive correlation between the addition of all stochastic couplings $v_i$ and the difference in the absorption redistribution $\Delta D_T^{full}-\sum_i\Delta D_i$  between the full system and all harvesting-target pairs. 3000 stochastic realisations made of parallel transition dipoles $\vec d_T$ and $\vec d_i$}\label{additivity}
\end{figure}

Another physically different system is the one we consider in this paper,  namely, many excitonic harvesting transitions which couple to a single target excited state. In this situation, if we consider again that all harvesting transitions have an energy $-\Delta E$ with respect to the target transition, and they all couple to this single transition with the coupling strengths $v_i$, the hamiltonian $H=\sum H_i$ is the addition of all pairwise Hamiltonians, however, spanning a $N+1$ excited state Hilbert space.  This case results in a  dipole strength $D_T^{full}$ for the target-like transition, with a dipole strength that changes by a relative amount $\Delta D_T^{full}=(|D_T^{full}|^2-|d_T|^2)/|d_T|^2$.
In order to compare the output from both situations, we considered $N$ harvesting transitions, proceed to generate  stochastic values $v_i$ and calculated $\sum_i^N \Delta D_i$ and  $\Delta D_T^{full}$, with events presented in Fig. \ref{additivity}. For the regime $\Delta E\gg v_i$, both  procedures result in a very similar absorption redistribution $\sum_i \Delta D_i\simeq D_T^{full}$, as can be deduced by the overlap between the coordinates $\{\sum_i \Delta D_i,D_T^{full}\}$ of events, and the line that represents the equality $\sum_i \Delta D_i= D_T^{full}$ in {\bf A}.  It can also be observed that those realisation with the greater values of absorption redistribution differ slightly from this additivity. This can be understood from Fig.\ref{additivity}{\bf B} where it can be observed that there is a positive correlation between the difference of the transition dipole strength redistribution $D_T^{full}-\sum_i \Delta D_i$  and the algebraic addition of all the couplings $v_i$ from each particular realisation. Accordingly, those realisations which present the highest couplings, will be those for which the additivity is no longer fulfilled, thereby suggesting that the additivity is a consequence of the perturbative regime $\Delta E\gg v_i$. Returning to Fig.\ref{additivity}{\bf A}, note that the dipole redistribution in all realisations observed is greater for the actual situation of $N$ harvester and a single target than for the one that stems for individual harvesting-target pairs, i.e.  $D_T^{full}>\sum_i \Delta D_i$, hence, the two level system description underestimates the absorption redistribution in the actual system, and therefore serves to illustrate the figures of merits there involved without artificially amplifying the effect.

\section{Modelling of optimal redirection upon incoherent excitation}
Given that  appreciable excitation with dim illumination from a spectrally broad quantised electromagnetic mode occurs in a time-scale longer than the correlations of the mode's field, its effect on the excitonic system can be captured perturbatively, 
by means of a Lindblad-type master equation with rates  obtained  by Fermi's golden rule. The excitation process occurs with a rate proportional to the dipole strength of the relevant states. For the full excitonic dephasing model, in the absence of the mode illumination, we obtain $\partial_t\rho_{\alpha{\displaystyle\prime},\alpha{\displaystyle\prime}}(t)=0$ which selects states $\alpha\prime$ as the preferred basis. A perturbative model yields to the illumination rate $\kappa=c| D_\alpha^{\displaystyle\prime}|^2$ and a correspondent Lindblad operator $O_{mode}=\sqrt{c} | D_\alpha^{\displaystyle\prime} |^2\ket{\alpha^{\displaystyle\prime}}\bra{0}$ for a constant $c\ll 1$. 
\begin{eqnarray}
\partial_t\rho_{00}=-c I \rho_{00}\nonumber\\
\partial_t\rho_{\alpha{\displaystyle\prime},\alpha{\displaystyle\prime}}=c| D_{\alpha}^{\displaystyle\prime}|^2 \rho_{00}
\end{eqnarray}
with solutions $\rho_{\alpha{\displaystyle\prime},\alpha{\displaystyle\prime}}=| D_\alpha^{\displaystyle\prime}|^2(1-e^{-cI t})/I$. Accordingly, the density operator in the excited state manifold must respect the ratio of these populations, fulfilled by
\begin{eqnarray}
 \rho(t^*)=\sum_{\alpha{\displaystyle\prime}} |D_\alpha^{\displaystyle\prime}|^2\ket{\alpha^{\displaystyle\prime}}\bra{\alpha^{\displaystyle\prime}}/I.\label{all_rho}
\end{eqnarray}

Hence, upon broadband excitation over the entire absorbing range and with protected RC-harvesting pigments electronic coherence encoded within $\alpha^{\displaystyle{\prime}}$ states, after the coherent population redirection occurs, $\rho(t^*)$ follows
\begin{align}
\rho(t^*)   &= \frac{1}{I}
          \bigg(|D_{P870}^{\displaystyle \prime}|^2\ket{P870^{\displaystyle \prime}}\bra{P870^{\displaystyle \prime}} +\nonumber\\
          & \qquad +|D_{B880}^{\displaystyle \prime}|^2\ket{B880^{\displaystyle \prime}}\bra{B880^{\displaystyle \prime}} \bigg).
\end{align}
 where we just took into account a single state in the RC and a single state from the LH unit, in a approximation that rests in the approximate additivity of the absorption redistribution from many LH states to a single RC state.  The population in this single special pair state, to be referred as target state $\ket{T}=\ket{P870}$ is calculated, using the states \ref{eq:app_Pwave} and  \ref{eq:app_Bwave}, with equation (3) of the main text
\begin{eqnarray}
P_T(t^*) &= \mbox{Tr} \left\{ \rho(t^*) \ket{T}\bra{T} \, \right\} \label{eq:Ppop_abs}\\
          &=\frac{1}{4 I} \bigg\{ |{D}_{T}|^2(3+\cos4\theta)+ \nonumber\\
          & \qquad  +2\Re\left\{{D}_{T}^*\cdot{D}_{LH}\right\}\sin4\theta+ \nonumber\\
          & \qquad  +|{D}_{LH}|^2(1-\cos4\theta) \bigg\} \label{eq:rho0_P}
\end{eqnarray}
where Tr is the trace operation and $LH$ stands for all quantities associated to the chosen B880 state of LH1. The above expression corresponds to equation (4) in the main text. Based on this expression and the fact that our numerical optimisation routines, to be discussed in further detail in section \ref{optimisation}, set configurations  with two bright states for optimal increase in dipole strength of a target pigment, a two level system is appropriate to explore the optimal energy landscape for the coherent population redirection, with the aim of elucidating design principles for artificial harvesting devices. For this case, we consider $\Re\{D_T^*\cdot D_{LH}\}=D_T D_{LH}$, and the energy landscape $\Delta E$ can be obtained by the optimal angle $\theta$ according to
\begin{align}
\frac{dP_T}{d\theta} \equiv 0 \rightarrow \tan 4\theta = \frac{2{\cal R}\{  D_T^*\cdot  D_{LH}\}}{|D_T|^2 - |D_{LH}|^2}
\end{align}
which results in equation (5) in the main text
\begin{eqnarray}\label{eq4}
P_T^{max} = \frac{D_{LH}^2+2D_T^2}{2 I}.
\end{eqnarray}
which displays a higher population redistribution than $P_T^{res} = (D_{LH}^2+D_T^2)/2 I$ whenever target and LH transitions are resonant $\Delta E \rightarrow 0$. Energy mismatch is thereby required for exploiting this coherent effect.

\begin{figure*}
\includegraphics[width=.49\columnwidth]{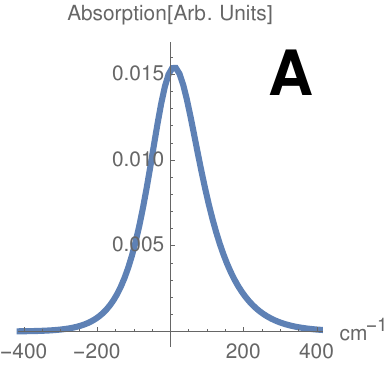}
\includegraphics[width=.49\columnwidth]{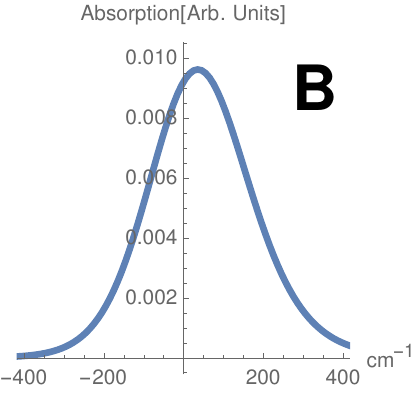}
\includegraphics[width=.49\columnwidth]{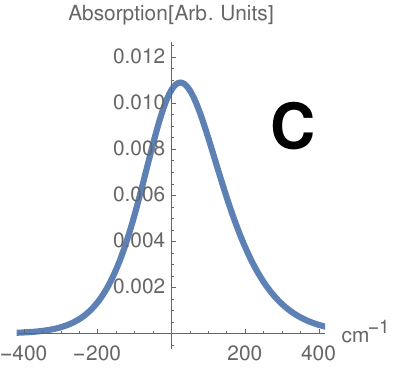}\\
\includegraphics[width=.5\columnwidth]{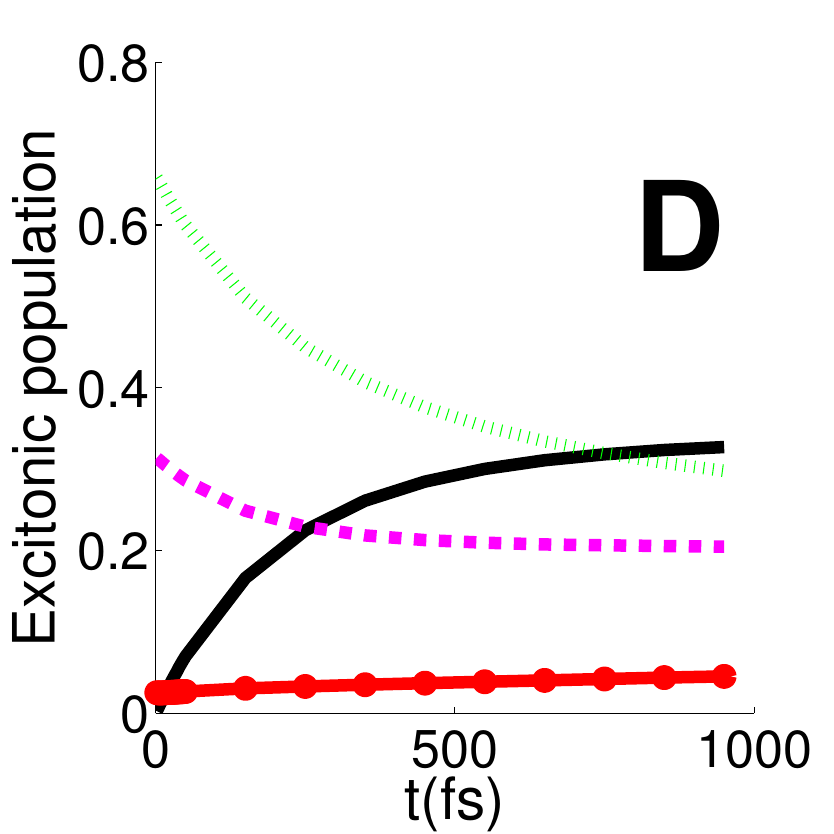}
\includegraphics[width=.5\columnwidth]{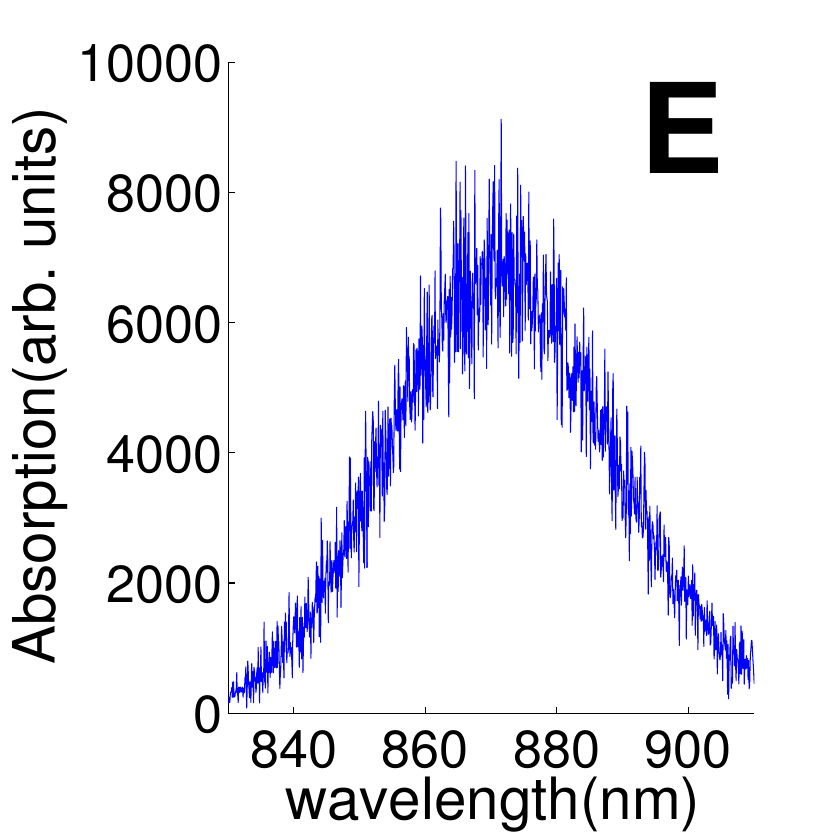}
\includegraphics[width=.49\columnwidth]{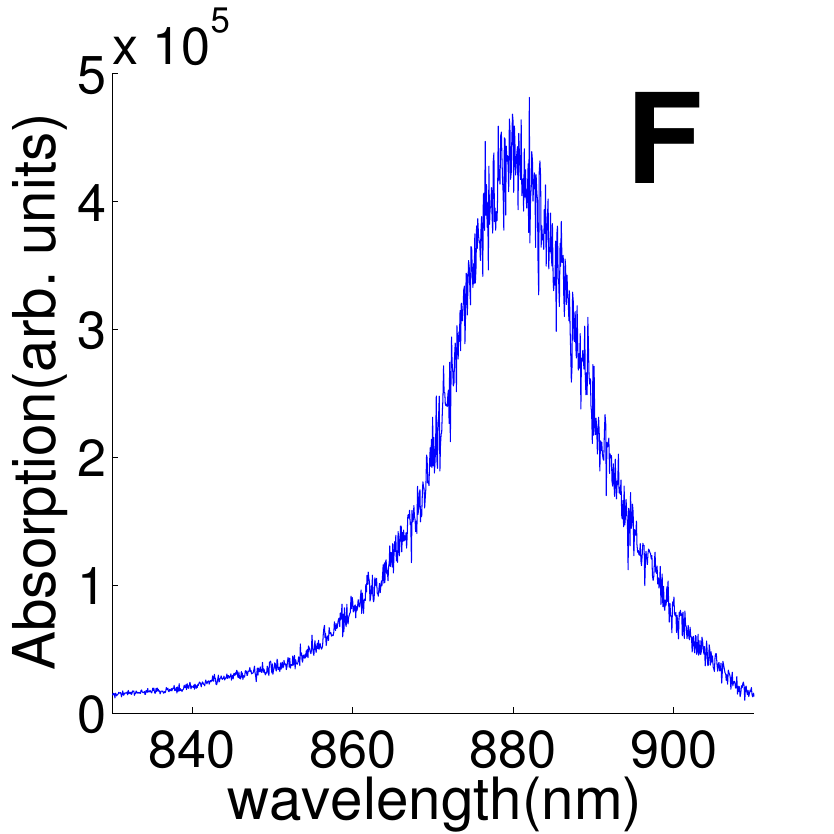}
\caption{Homogeneous spectrum resulting from line-shape theory analysis. {\bf A} Result with just the pure dephasing contribution being accounted for (parameters \cite{Jimenez_JPCB1997}: $\tau_c= 60$ fs, $\sqrt{\langle\Delta\omega_1^2\rangle}=100$ cm$^{-1}$), and {\bf B} shows both dephasing and intra-ring transfer contributions (transfer parameters \cite{Jimenez_JPCB1997}: $\tau_t= 90$ fs, $\sqrt{\langle\Delta\omega_2^2\rangle}=160$ cm$^{-1}$). {\bf C} Presents the homogeneous spectrum for the RC (parameters \cite{Jordanides_2001JPCB}: $\tau_c= 60$ fs, $\sqrt{\langle\Delta\omega_1^2\rangle}=130$. {\bf D} Presents populations in a single realisation of the first four  -in energy ascending order- excitonic states $\alpha=1,2,3,4$ in continuous black,  dotted-green, dashed-magenta and bullets-continuous-red, respectively.  Inhomogeneously broadened stick spectra of  P870  and  B880 in {\bf E} and {\bf F} present standard deviations of 195 cm$^{-1}$ and 110 cm$^{-1}$, respectively. The stick spectra is the result of  weighted histograms of the eigenstates dipole strength at the respective frequency. $5\times 10^4$ realisations of diagonal and non-diagonal noise in the ring and diagonal noise in the P pigments, as described in Tables S1-S2.}\label{lineshape}
\end{figure*}

\section{Estimation of parameters under physiological conditions}
\subsection{LH1 nearest neighbours coupling strength}
Nearest neighbour couplings $Q_1$ and $Q_2$ and their interplay with the static disorder, determines the length of excitons, if no dephasing is accounted for. For the LH1, it has been shown that inhomogeneities of the same magnitude as the nearest neighbour coupling strengths (300 and 233 cm$^{-1}$ in LH1 intra- and inter-dimer couplings), result in a delocalization length of excitons $\alpha$ of about 3-4 pigments, compatible with the number of pigments involved in the fluorescence superradiance observed \cite{vanGrondelle_1997JPCB} and circular dichroism measurements \cite{VanGrondelle_JP2006}. Since our partial excitonic model is based on dephasing of states $O_\alpha$, it  does not dynamically degrade the delocalisation length over either RC or LH1 excitons given by this compromise, thereby,  it  fulfils the most relevant observations to describe absorption spectra and dynamics, we are aware of to date.

\subsection{Dephasing and relaxation rates}

 Access to the parameters that characterise individual contributions to the full spectral broadening was made through three pulse photon echo peak shift (3PEPS) observations which determined decay time constants and lineshape coupling strengths of the electronic energy gap correlation function. Accordingly, the most relevant broadening contributions in the LH1 were ascribed to pure dephasing and intra-ring incoherent transfer dynamics \cite{Jimenez_JPCB1997}, from which we constructed the homogeneous spectrum associated to these processes based on the line-shape theory and parameters utilised to analyse such data, as follows.

Dissipation from excited to ground states with a rate of 1/600 ps as was experimentally observed \cite{Freiberg92}. The time correlation function of the electronic transition frequency $M_i(t)$ are the elements of direct access from three pulse photon echo peak shift (3PEPS), which register in general  photosynthetic complexes, a contribution with a gaussian functional dependence, $M_1(t)=e^{-(t/\tau_c)^2}$ that has been identified with pure dephasing \cite{Jimenez_JPCB1997}. The limiting case of fast fluctuations $1/\tau_c\gg\sqrt{\langle \Delta\omega^2\rangle}$ results in the Markovian scenario, i.e., with  Lorentzian lineshapes. For the case of LH1, an additional exponential component $M_2(t)=e^{-(t/\tau_t)}$ was identified, and ascribed to intra-ring relaxation \cite{Jimenez_JPCB1997}. Following the formalism of lineshape based in a cumulant expansion  \cite{Kubo_1962JPSJ}  we obtained homogeneously broadened spectra shown in Fig.\ref{lineshape}.  The dephasing contribution leads to the lineshape presented in A, with  a full width half maximum (FWHM) of 172 cm$^{-1}$. The addition of the intra-ring excitonic relaxation contribution results in an homogeneous spectrum presented in  B, with a FWHM of 294 cm$^{-1}$. For the case of the RC a similar treatment, however with  experimental 3PEPS data from accessory pigments \cite{Flemming_1998JPCB} and not from the P870 transition. In this case, since no incoherent transfer component was observed, just  dephasing results in a homogeneous contribution in Fig.\ref{lineshape} C, with a FWHM of 254 cm$^{-1}$. The FWHM from the dephasing contribution were set equal to $\gamma_{\alpha}^{\displaystyle\prime}$ or $\gamma_{\alpha}$  for the respective full exciton or partial exciton models, while the full FWHM were set equal to $\gamma_i$ in the site dephasing model.

The values for the inter-excitonic relaxation rates $\gamma_{\alpha,\beta}\propto J(\omega_\alpha-\omega_\beta)$, are proportional to the spectral density of $J(\omega_\alpha-\omega_\beta)$ with a functional form determined in B877 monomeric units \cite{Renger2006}. The proportionality constant was chosen such that the equilibration of the ring system is achieved within 400 fs -as observed through equilibration of the B880 in pump-dump-probe experiments \cite{Cohen_2011BioJ}-, with excitonic dynamics presented in Fig.\ref{lineshape} D. The initial condition for the simulations shown in Fig.2 in the main text corresponds to the ground state $\rho(0)=\ket{0}\bra{0}$. The negligible value of the function $J(\omega_\alpha-\omega_\beta)$  for the splitting between $P$ states of $\simeq$ 1000 cm$^{-1}$, supports a minor contribution from relaxation in the P870 band.

\subsection{Inhomogeneous broadening mechanisms}\label{inh_broad}
The homogeneous broadening complements the  inhomogeneities present in the ensemble, in order to adjust the spectral broadening  in the measurement of macroscopic samples.
  
For the case of the LH1, these inhomogeneities are thought to arise --as introduced in our model-- from fluctuations in pigment energies $\omega_i$ (standard deviation 300 cm$^{-1}$) and nearest neighbour excitonic couplings $J_{i,i\pm1}$ (standard deviation 180 cm$^{-1}$),  \cite{Timpmann_2005CPL} while no evidence has been found so far for important excitonic coupling disorder in the RC. Hence only variations of the $P$ pigment energies are performed (standard deviation 310 cm$^{-1}$). These variations produce an effective standard deviations  of 195 cm$^{-1}$ and 110 cm$^{-1}$ highlighted in Fig.\ref{lineshape}) E and F for the P870 and B880 resonances, respectively.

\section{Determination of RC-like and LH1-like spectra}\label{determ_like}
For each realisation of inhomogeneous noise, a diagonalization of the electronic Hamiltonian  equation (1) in the main text results in eigenstates $\ket{\alpha^{\displaystyle \prime}}$. The RC-like (LH1-like) states correspond  to those states $\alpha^{\displaystyle\prime}$, labeled as $\alpha_{RC}^{\displaystyle\prime}$ or $\alpha_{LH}^{\displaystyle\prime}$,  which mainly delocalize over the RC or the LH1 pigments, respectively. The dipole operator associated with these states $ D_{RC}=\sum_{\alpha\prime\epsilon RC}D_{\alpha^\prime}\ket{0}\bra{\alpha^{\displaystyle\prime}}+h.c$
and $D_{LH}=\sum_{\alpha\prime\epsilon LH}D_{\alpha\displaystyle\prime}\ket{0}\bra{\alpha^{\displaystyle\prime}}+h.c$, and its time evolution $ D_{RC}(t)$ and $ D_{LH}(t)$ are thereby the building blocks for the construction of RC-like and LH1-like spectra, presented in Fig.1 C-D in the main text. The scalar quantities $D_{RC}$ or $D_{LH}$ account for the component of the respective dipole moment vector along the direction of the incident electric field with a relative orientation changing among realisations. The rather low intensity illumination used in optical experiments with photosynthetic complexes, results in a stationary density operator $\rho_{ss}\approx \ket{0}\bra{0}$ which under the action of the, e.g., RC-like dipole moment operator, transforms into $\sum_{\alpha\prime\epsilon RC} D_{\alpha\prime}\ket{\alpha\prime}\bra{0}$. The Laplace-Fourier transform of the time evolution super-operator results in the super-operator $1/({\mathcal L}-i\omega)$ for a master equation $\partial_t\rho=\mathcal{L}\rho$, $\rho(t)=e^{\mathcal L t}\rho(0)$, as was performed in \cite{Plenio_JCP2013}. For the RC-like spectrum  we obtain $A(\omega)_{RC}=\mbox{ Re\{Tr}\{D_{RC}(i\omega-{\mathcal L})^{-1}D_{RC}\rho_{ss}\}\}$ for every single realisation, and then we average the full ensemble to result in the spectra shown in Fig.1 of the main text, or Fig.\ref {comparison_spectra} below.

\section{Approximate equivalence of microscopic dephasing models for calculation of absorption spectra}

\subsection{Equivalence based on the DDCF calculation}

\begin{figure*}
\includegraphics[width=.59\columnwidth]{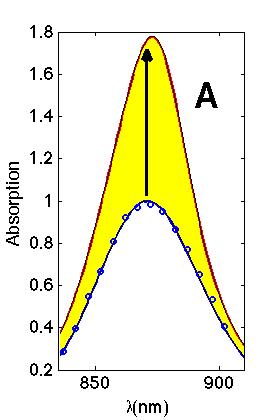}
\includegraphics[width=.59\columnwidth]{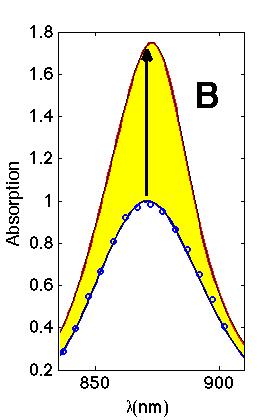}
\includegraphics[width=.59\columnwidth]{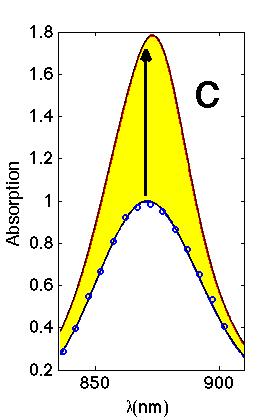}
\caption{RC (continuous blue theory, circles experiment) and RC-like (red) spectrum for {\bf A} site dephasing, {\bf B} partial exciton dephasing models and {\bf C} full core exciton dephasing models. Yellow area highlights the difference between RC-like and RC spectrum. In {\bf A} and {\bf C} the only mechanism of dephasing is site dephasing and full core excitons dephasing, with rates $\gamma_i$ and $\gamma_\alpha^{\displaystyle\prime}=254$ and $294$ cm$^{-1}$. Average from $3\times 10^4$ realisations of inhomogeneities.}\label{comparison_spectra}
\end{figure*}

Within a Markovian modelling, the pure local dephasing contribution is described by a Lindblad dephasing operator ${\cal L}_d\rho=\sum_i  O_i\rho O_i-\frac{1}{2}O_i^+O_i\rho-\frac{1}{2}\rho O_i^+O_i$, with $O_i=\sqrt{\gamma_i}\ket{i}\bra{i}$. The terms $O_i\rho O_i^+=\gamma_i\ket{i}\bra{i}\rho\ket{i}\bra{i}$  influence the pigments' excited state populations, i.e. $\bra{i}O_i\rho O_i^+\ket{i}\ne 0$. These term will influence excitonic coherences, i.e. $\bra{\alpha}O_i\rho O_i^+\ket{\beta}\ne 0$, but  will not contribute to the equations of motion of optical coherences since  $\bra{\alpha}O_i\rho O_i^+\ket{0}=\bra{i}O_i\rho O_i^+\ket{0}= 0$.
Hence, only the terms $O^+O\rho$ and $\rho O^+O$  influence the DDCF and absorption spectra lineshapes.

In what follows we will present the regimes where the absorption spectrum shows similar features for different dephasing models. The terms  $O^+O\rho$ and $\rho O^+O$ are written for the site basis dephasing 
$$\sum_i\gamma_d^i\ket{i}\bra{i}=\sum_i\gamma_d^{RC}\ket{i}\bra{i}+\sum_i\gamma_d^{LH}\ket{i}\bra{i}$$
where it has been assumed that the dephasing rates of pigments within the RC or within the LH1 structures  present equal dephasing rates. 

Now, if the above expression is transformed to the partial excitonic basis we obtain
\begin{eqnarray} 
\sum_i\gamma_d^i\ket{i}\bra{i}=\sum_{i,\alpha,\beta}\gamma_d^i c^{\alpha}_i c^{\beta}_i\ket{\alpha}\bra{\beta}, \label{site_deph}
\end{eqnarray}
where for notational convenience we have assumed $c_i^\alpha=(c_i^\alpha)^*$. 
This expression can be split into contributions from the overlap of excitons delocalised over the LH1 or delocalised over the RC. Since the partial delocalisation over RC or LH units results in cross terms $c_i^{\alpha\epsilon RC} c_i^{\beta\epsilon LH}= 0$, then 
 $$\sum_i\gamma_d^i\ket{i}\bra{i}= \gamma_d^{RC}\sum_{i,\alpha\epsilon RC,\beta\epsilon RC} c^{\alpha}_i c^{\beta}_i\ket{\alpha}\bra{\beta}$$ $$+\gamma_d^{LH}\sum_{i,\alpha\epsilon LH,\beta\epsilon LH} c^{\alpha}_i c^{\beta}_i\ket{\alpha}\bra{\beta}.$$
The same condition of partial delocalisation restricts excitons $\alpha$ and $\beta$ to either RC or LH pigments and the partial closure relation $\sum_{i\epsilon RC} c_{\alpha}^i c_{\beta}^i=\sum_{i\epsilon LH} c_{\alpha}^i c_{\beta}^i\approx \delta_{\alpha,\beta}$ is approximately fulfilled. Therefore from equation (\ref{site_deph}) we obtain

\begin{eqnarray}
\sum_{i\epsilon LH}\gamma_d^{LH}\ket{i}\bra{i}+\sum_{i\epsilon RC}\gamma_d^{RC}\ket{i}\bra{i}&=& \sum_{\alpha\epsilon LH} \gamma_d^{LH}\ket{\alpha}\bra{\alpha}+\sum_{\alpha\epsilon RC} \gamma_d^{RC}\ket{\alpha}\bra{\alpha}.\nonumber\\
&&\label{equality}
\end{eqnarray}
This result shows that the microscopic model of site or partial excitons dephasing are {\it approximately equivalent regarding the calculation of DDCF}.  Nevertheless, the partial excitonic dephasing is just a contribution to the total homogeneous broadening in this model, as we have also considered intra-ring excitonic decay. Given that the relaxation is smaller than the dephasing component, it is expected for the relaxation not to change dramatically the characteristics of the absorption spectra. The demonstration for the approximate equivalence for absorption spectra of site and partial excitons dephasing models is presented in Fig.\ref{comparison_spectra}. In A, RC and RC-like spectra are shown for the site dephasing model, and in B  the  analogue quantities for the partial excitons dephasing model are presented. For the definition of RC-like spectra, see section \ref{determ_like}. The predictions of both models are very similar indeed, regarding  increase an increase of the P870 cross section of 61\% and 59\% for the site  and the partial excitonic dephasing plus intra-ring relaxation models, respectively. The absorption redistribution to the RC is similar in the full core complex dephasing model as presented in C.  The site dephasing written in the full core complex eigenstate basis results in
$$\sum_i\gamma_d^i\ket{i}\bra{i}= \gamma_d^{RC}\sum_{i,\alpha{\displaystyle\prime}\epsilon RC,\beta{\displaystyle\prime\epsilon} RC} c^{\alpha\displaystyle\prime}_i c^{\beta\displaystyle\prime}_i\ket{\alpha^{\displaystyle\prime}}\bra{\beta^{\displaystyle\prime}}$$ $$+\gamma_d^{LH}\sum_{i,\alpha{\displaystyle\prime}\epsilon RC,\beta{\displaystyle\prime}\epsilon LH} c^{\alpha\displaystyle\prime}_i c^{\beta\displaystyle\prime}_i\ket{\alpha^{\displaystyle\prime}}\bra{\beta^{\displaystyle\prime}}$$
where now the approximation consists on having almost zero overlap between RC-like and LH1-like states, i.e.  $c_{i}^{\alpha\prime \epsilon RC} c_{i}^{\beta\prime\epsilon LH}\approx 0$ while the closure relation is exact $\sum_{i} c_{\alpha\prime}^i c_{\beta\prime}^i=\delta_{\alpha\prime,\beta\prime}$, then we obtain expression \ref{equality} however with labels $\alpha,\beta$ exchanged to $\alpha^{\displaystyle\prime},\beta^{\displaystyle\prime}$. Notice that this just commented approximation is well founded when the delocalisation over both RC and LH pigments is minor which is well fulfilled by the gross set of realisations where the mixing angle will be small, namely $\Delta E/V\ll 1$, a condition fulfilled in average for all the LH-RC interactions we are aware of. 

\subsection{Lineshapes dressing procedure}

The model of  dephasing between full core excitons with Lindblad operators $O_\alpha^{\displaystyle\prime}=\sqrt{\gamma_\alpha^{\displaystyle\prime} }\ket{\alpha^{\displaystyle\prime}}
\bra{\alpha^{\displaystyle\prime}}$ results in equations of motion for the optical coherences 
\begin{equation}
\partial_t\rho_{\alpha{\displaystyle\prime},0}=-\frac{\gamma_\alpha^{\displaystyle\prime}}{2}\rho_{\alpha{\displaystyle\prime},0}.
\end{equation}
This evolution results in the full protection of the coherence between RC and LH pigments of a given state $\alpha^{\displaystyle\prime}$ although it decreases the coherence of it with respect to other $\ket{\beta^{\displaystyle\prime}}$ states. With this model, an initial state $\rho=\ket{\alpha^{\displaystyle\prime}}\bra{\alpha^{\displaystyle\prime}}$ will not deteriorate its coherence properties between pigments in time. This model sets a Lorentzian homogeneous spectra profile  with FWHM $2\gamma_\alpha^{\displaystyle\prime}$ and the calculation of the spectrum becomes equivalent to dressing the stick spectra of realisation energies $\omega_{\alpha}^{\displaystyle\prime}$, by means of a Lorentzian function with FWHM $2\gamma^{\displaystyle\prime}_\alpha$. The result is presented in
Fig.\ref{comparison_spectra}C, with a very close resemblance to the site dephasing model, which nevertheless does not protect the LH-RC electronic coherence at all. Therefore, the procedure of dressing stochastic realisations of the Hamiltonian by means of Lorentzian functions, is adequate in order to evaluate the possibility to detect the influence of moderate couplings into the dynamics of excitonic systems,  in a manner that is robust to the specific microscopic dephasing model.

\section{Model for dynamics under incoherent illumination}
In order to simulate the dynamics of the core complex when subject to incoherent light we use the full core complex excitonic Hamiltonian  equation (1) in the main text which couples to a bosonic mode in contact with a Markovian reservoir at temperature $T=5000 K$ to model irradiation by sunlight.  The Hamiltonian for this model 
\begin{equation}
H_{full}={\mathcal H}+\hbar \Omega a^+a+\sum_i g\vec d_i\cdot \hat E (a^+\ket{0}\bra{i}+a\ket{i}\bra{0})
\end{equation}
stand under  the usual rotating wave approximation since the  oscillating components of the field at double the optical frequency (with time-scale$\approx$ 1 fs) are very fast compared with the coherent population dynamics occurring, however, over a time-scale given by the inverse of the homogeneous spectral linewidth ($\approx$ 70 fs). The electromagnetic mode is characterised by bosonic creation and annihilation operators $a^+$ and $a$, whose frequency $\Omega=(\omega_{B880}+\omega_{P870})/2\approx  2\pi/875$ nm is set tuned to the average frequency of the two bands. The exact tuning of the mode is, as will be shortly seen, not as important.

The dynamics of the electromagnetic mode in contact with a thermal Markovian reservoir is modelled by Lindblad dynamics. The Lindbladian operators are given by $O_\uparrow=\sqrt{\Gamma(n+1)}a^+$ and  $O_\downarrow=\sqrt{\Gamma n}\, a$ where $n$ represent the mean number of bosons at the frequency $\Omega$ from  a thermal source at temperature 5000 K. In order to guarantee that the mode describes illumination from an incoherent source, we use $g=4\times 10^{-7}$ cm$^{-1}$/Debye for a value that stems for a Rabi frequency of a monochromatic field with intensity of 0.1 W/m$^2$, greater than the usual growth intensity for {\it R. rubrum} but common to other purple bacteria environment. The value of $\Gamma=10^4$ cm$^{-1}$ and was chosen much greater than the energy difference between the bands $\Delta E$ and gauged such that further increase of its value, did not change the presented results.

\section{Predicted effects in the absorption spectra of Higher Plants}
\begin{figure}[!t]
\includegraphics[width=.7\columnwidth]{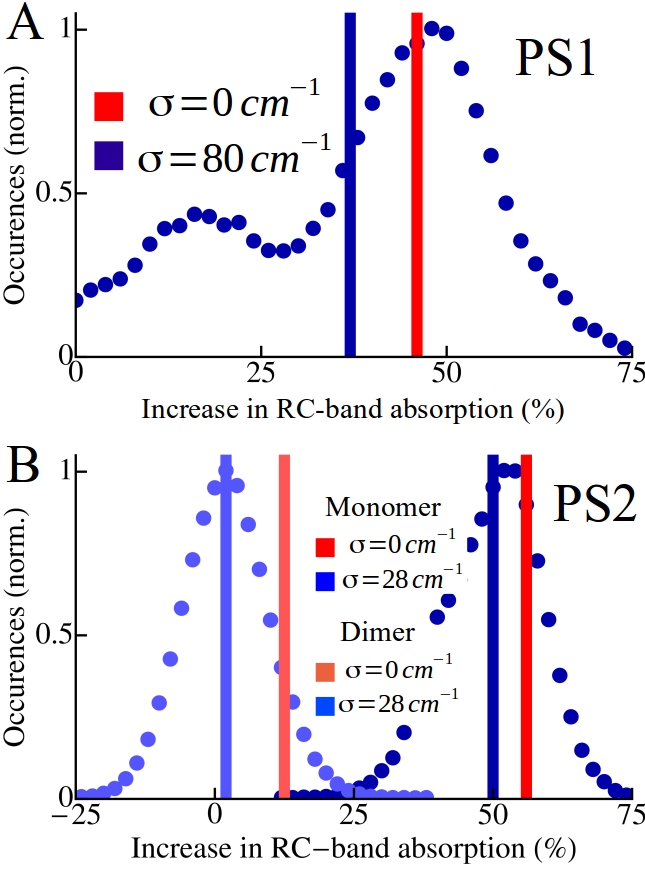}
\caption{Histograms of absorption cross section redistribution from antenna to $P$ pigments bands in PS1 ({\bf A}) and PS2 ({\bf B}) from realisations  of site energies Gaussian disorder. The average change of $P$ absorption cross section in PS1 shifts from $+46\%$ with no disorder to $+35\%$ with $80$ cm$^{-1}$ inhomogeneous broadening. In the PS2 dimer, the absorption redistribution shifts from $+11\%$ with no disorder to $+2\%$, with $28$ cm$^{-1}$ inhomogeneous broadening, and the PS2 monomer shifts from $+54\%$ to $+50\%$, with the same inhomogeneous broadening.}\label{fig:SIPS}
\end{figure}

Photosystem 1 (PS1) and photosystem 2 (PS2) are the main light-harvesting photounits in higher plants \cite{Ort96}. To 
explore the effects of extended delocalisation in these systems, we examine Hamiltonians of the form
\begin{align}
\mathscr{H}=	\begin{pmatrix}
		\mathscr{H}_{RC} & V \\
		V & \mathscr{H}_{LH} \\
	\end{pmatrix}
\end{align}
where $\mathscr{H}_{RC}$ is the Hamiltonian of the full RC alone and $\mathscr{H}_{LH}$ is the Hamiltonian of the antenna complexes. The coupling between RC and LH pigments is captured by the matrix $V$. In this simplified calculation, site energies are taken from the literature and couplings between pigments are calculated in the point-dipole approximation using $d=4.48$ D, $\kappa=1$ and $d=4.4$ D, $\kappa=1.5$ for PS1 and PS2, respectively \cite{Damjanovic_02,Raszewksi_08,Raszewksi_05}. $\kappa$ is the relative permitivity. Transition dipoles, $\vec{d}_i$, are oriented along the direction connecting the Nitrogen atoms of $B$ and $D$ porphyrin rings of each chlorophyll using the atomic coordinates taken from the X-ray structures (PDB accession codes 1JB0 and 3WU2), and rotated $15\degree{}$ towards $N_C$ in the case of PS2 \cite{Weiss72,Raszewksi_08}. Fluctuations in the pigment energies, drawn from a Gaussian distribution, capture spectral inhomogeneities, which are known to fluctuate with temperature. We take Gaussian standard deviations of $\sigma$ = $80$ cm$^{-1}$ (fit from low temperature spectra) and $\sigma$ = $28$ cm$^{-1}$ (fit from room temperature spectra) for PS1 and PS2, respectively\cite{Yin_07,Holzwarth_96}. For each realisation of the Hamiltonian, $\mathscr{H}$, we find  the dipole moments of the RC-like exciton states, $\vec D_\alpha^{\displaystyle\prime}$ and proceed with the histogram of the ratio  $\sum_{\alpha\prime\epsilon \mbox{\tiny{RC-like}}}|\vec D_\alpha^{\displaystyle\prime}|^2/|\sum_{i\epsilon RC} |\vec d_i|^2$ which accounts to the ratio between the RC dipole strength for $V\neq0$ (coherently coupled RC and antennae) and the analogue quantity for $V=0$ (uncoupled RC and antennae).
As summarised in Fig.~\ref{fig:SIPS}, these calculations predict an increase in the excitation of $P$ pigments due to excitonic delocalisation over antenna chlorophylls which is robust to, albeit reduced by, static disorder.

\begin{figure}
\includegraphics[width=1\columnwidth]{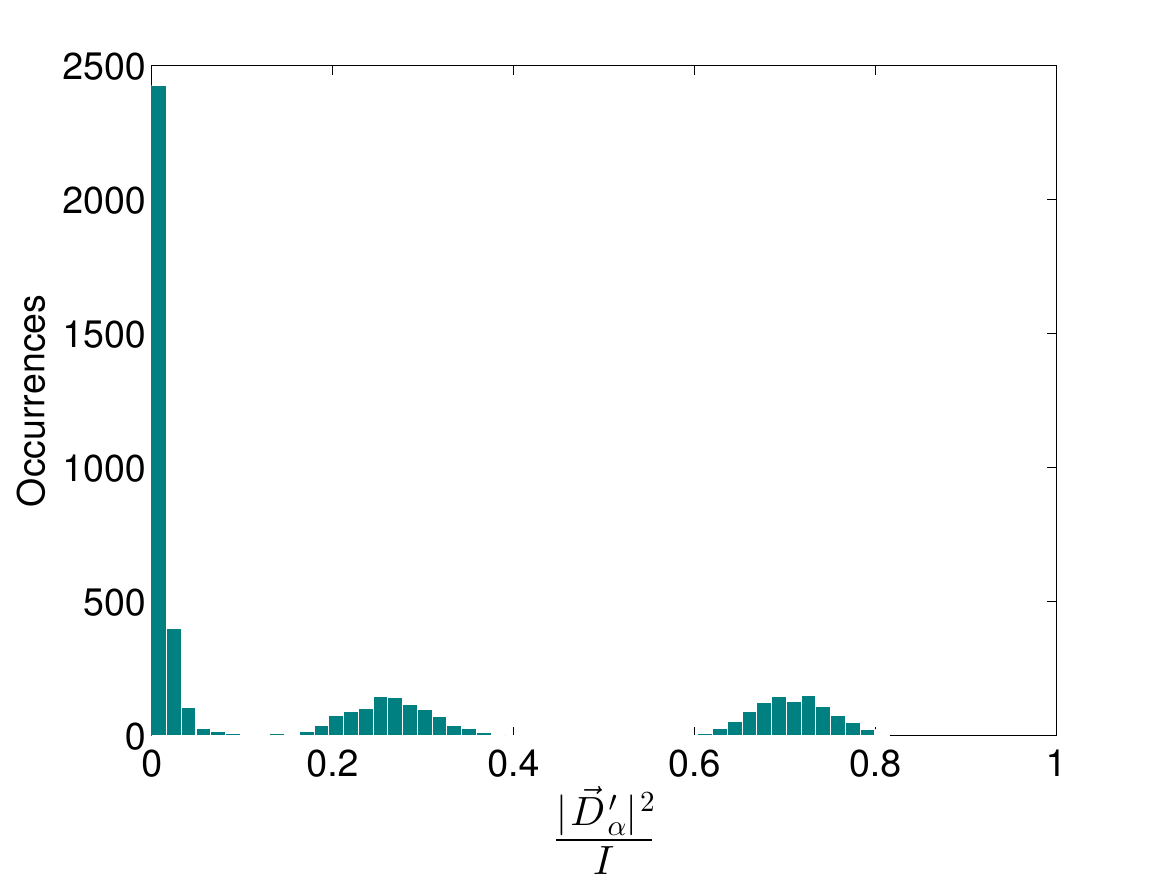}
\caption{Histogram of dipole strengths for optimal configurations in regards to the greatest redirection of population to the target pigment, with $N+1=5$ dipoles. Only two states $\alpha^{\displaystyle\prime}$ present appreciable dipole strength, supported by the fact that these two sets comprise one fifth of the total occurrences. 1000 optimal configurations.}\label{optimisation}
\end{figure}

\section{Optimisation procedure for population redirection in artificial devices}\label{optimisation}

The parameters that were varied in order to optimise the coherent population redirection were the coordinates of harvesting and target units, the direction $\hat{d}_i$ of their transition dipoles, and the energies of individual pigments $\omega_i$. Based on an initial test, we noticed that configurations where all pigments lie on a plane converged more rapidly to optimal performance. Given that the interplay between positions --that result in respective excitonic coupling-- and pigment energies $\omega_i$, determine the dipole strengths $|\vec D_\alpha^{\displaystyle\prime}|^2$, we chose only to vary the positions of pigments and transition dipoles orientation.  We therefore fixed the energies of all pigments to the same value.
For simplicity, a single target pigment was chosen and its population based on eqs. (\ref{all_rho}) and (\ref{eq:Ppop_abs})  was calculated.  Realisations proceed by stochastic drifts of the pigments' coordinates and/or dipole moments direction whenever the target pigment has improved its population after coherent redirection, i.e. increase in $P_T(t^*)$ of equation (\ref{eq:Ppop_abs}).

Besides the advantage of planar configurations, a simple inspection of the direction from the transition dipoles showed that optimal population redirection arose from pigments with close to parallel transition dipoles. As a result, random configurations of pigments which lie in a common plane and with parallel transition dipoles were considered, and showed faster convergence to optimal values, as Table \ref{table_opt} reflects. The positions of the pigments in the plane, however, did not exhibit a typical arrangement for the optimal configurations. Though, inspection of  the  dipole strengths $|\vec D_\alpha^{\displaystyle \prime}|^2$ from these optimal configurations, show in Fig.~\ref{optimisation} that only two optical transitions develop.  We found that the two bright states had most population spread over either the target or the harvesting pigment. Accordingly,  removal of the target pigment from these optimal assemblies results in  only one optical transition for the remaining $N$ harvesting pigments, as presented in Fig.~3A in the main text. Notice that none of the optimal configurations develops a single bright state with the complete dipole strength $I=\sum _i^{N+1} |\vec d_i|$.

\section{Supplementary Tables}
\begin{table}[h!]
\caption{Parameters for the  RC of {\it R. rubrum} which capture the absorption spectrum of Fig. 1 in the main text.}\label{table_finalRC_dark}
\begin{center}
\begin{tabular}{|c|c|p{4cm}|}
\hline
Parameter              &   Value                     &       Description                 \\ \hline
$\omega_P$     &       11\, 900                & Energy of $P$ pigments	\\ \hline
$J_{P}$ 		& 	500 cm $^{-1}$ 	& $P$ excitonic coupling\\	 \hline
$d_{P}$                &  6.3 D                      & dipole moment for $Q_y$ transition of P BChls    \\ \hline
$\sigma_{P}$           & 310 cm$^{-1}$                & magnitude of  P  pigments energy disorder   \\ \hline
$\gamma_{P}$     & 195 cm$^{-1}$     & dephasing rate P pigments/excitonic transition  \\ 
\hline
\end{tabular}
\end{center}
\end{table}

\begin{table}[h!]
\caption{Parameters for the LH1 of {\it R. rubrum} which capture the absorption spectrum of Fig. 
1 in the main text.}\label{table_finalLH1}
\centering
\begin{tabular}{|c|c|p{4cm}|}
\hline
Parameter                       &   Value                       &       Description                 \\ \hline
$\gamma$                        & $10.06^{\circ}$               & intra-dimer angle                 \\ \hline
$\Delta\gamma_1,\Delta\gamma_2$ & $19.9^{\circ},17.6^{\circ}$   & dipole tilt away from tangent     \\ \hline
$\phi_1,\phi_2$                 & $6.24^{\circ}$, $4.81^{\circ}$& dipole tilt out of membrane plane \\ \hline
$r_1,r_2$                       & 46.51 $\AA$, 47.27 $\AA$      & radius of $\alpha$, $\beta$ rings \\ \hline
$z$                             & 0.63 $\AA$                    & vertical displacement among $\alpha$ and $\beta$ chromophores  \\ \hline
$d_{LH1}$                       & 9.8 D                   & dipole moment for $Q_y$ transition of LH1    \\ \hline
$\kappa$                        & 1.4                       & relative permittivity             \\ \hline
$\epsilon_\alpha,\epsilon_\beta$& 12646,12656 cm$^{-1}$   & chromophore energies  of $\alpha$ and $\beta$ site      \\ \hline
$Q_1, Q_2$                      & 300, 233 cm$^{-1}$      & intra- and inter-dimer excitonic coupling   \\ \hline
$\sigma_{J}$                    & 180 cm$^{-1}$            & magnitude of nearest neighbours coupling strength disorder\\ \hline
$\sigma_{LH1}$                  & 300 cm$^{-1}$            & magnitude of LH1 pigments energy disorder  \\ \hline
$\gamma_{LH1} $                 & 254  cm$^{-1}$               & dephasing rate LH1 pigments/excitonic transitions \\
 \hline
\end{tabular}
\end{table}

\begin{table}[h!]
\caption{ Values for the maximum population of the target pigment $P_T$ obtained by numerical optimisation. For each number $N$ of dipoles,   $10^6$ optimisation steps  were performed. In this data, the variations of just positions was made with fixed  parallel transition dipoles, which were parallel to the plane where the pigments lie.}\label{table_opt}
\begin{center}
\begin{tabular}{|c |c |c |c |c |c |}
\cline{2-6}
 \multicolumn{1}{c|}{} & \multicolumn{5}{c|}{$P_T$} \\
\hline
Variations & $N+1=3$ & $N+1=4$  &$N+1=5$ & $N+1=6$ &$N+1=7$ \\
\hline
position, dipoles      &   0.665 & 0.601 & 0.564 & 0.522 &0.493\\ \hline
positions	    & 	  0.667 	&  0.620 & 0.597 & 0.573 &0.522\\ \hline
Eq.(\ref{eq4})    & 0.667 & 0.625 &0.600& 0.583& 0.571\\ \hline
\end{tabular}
\end{center}
\end{table}



\begin{thebibliography}{33}
\expandafter\ifx\csname natexlab\endcsname\relax\def\natexlab#1{#1}\fi
\expandafter\ifx\csname bibnamefont\endcsname\relax
  \def\bibnamefont#1{#1}\fi
\expandafter\ifx\csname bibfnamefont\endcsname\relax
  \def\bibfnamefont#1{#1}\fi
\expandafter\ifx\csname citenamefont\endcsname\relax
  \def\citenamefont#1{#1}\fi
\expandafter\ifx\csname url\endcsname\relax
  \def\url#1{\texttt{#1}}\fi
\expandafter\ifx\csname urlprefix\endcsname\relax\def\urlprefix{URL }\fi
\providecommand{\bibinfo}[2]{#2}
\providecommand{\eprint}[2][]{\url{#2}}

\bibitem[{\citenamefont{Jordanides et~al.}(2001)\citenamefont{Jordanides,
  Scholes, and Fleming}}]{Jordanides_2001JPCB}
\bibinfo{author}{\bibfnamefont{X.~J.} \bibnamefont{Jordanides}},
  \bibinfo{author}{\bibfnamefont{G.~D.} \bibnamefont{Scholes}},
  \bibnamefont{and} \bibinfo{author}{\bibfnamefont{G.~R.}
  \bibnamefont{Fleming}}, \bibinfo{journal}{J. Phys. Chem. B}
  \textbf{\bibinfo{volume}{105}}, \bibinfo{pages}{1652} (\bibinfo{year}{2001}).

\bibitem[{\citenamefont{{Adolphs} and {Renger}}(2006)}]{Renger2006}
\bibinfo{author}{\bibfnamefont{J.}~\bibnamefont{{Adolphs}}} \bibnamefont{and}
  \bibinfo{author}{\bibfnamefont{T.}~\bibnamefont{{Renger}}},
  \bibinfo{journal}{Biophys. J.} \textbf{\bibinfo{volume}{91}},
  \bibinfo{pages}{2778} (\bibinfo{year}{2006}).

\bibitem[{\citenamefont{{Georgakopoulou}
  et~al.}(2006)\citenamefont{{Georgakopoulou}, {van Grondelle}, and {van der
  Zwan}}}]{VanGrondelle_JP2006}
\bibinfo{author}{\bibfnamefont{S.}~\bibnamefont{{Georgakopoulou}}},
  \bibinfo{author}{\bibfnamefont{R.}~\bibnamefont{{van Grondelle}}},
  \bibnamefont{and} \bibinfo{author}{\bibfnamefont{G.}~\bibnamefont{{van der
  Zwan}}}, \bibinfo{journal}{J. Phys. Chem. B} \textbf{\bibinfo{volume}{110}},
  \bibinfo{pages}{3344} (\bibinfo{year}{2006}).

\bibitem[{\citenamefont{{Timpmann} et~al.}(2005)\citenamefont{{Timpmann},
  {Trinkunas}, {Qian}, and {Hunter}}}]{Timpmann_2005CPL}
\bibinfo{author}{\bibfnamefont{K.}~\bibnamefont{{Timpmann}}},
  \bibinfo{author}{\bibfnamefont{G.}~\bibnamefont{{Trinkunas}}},
  \bibinfo{author}{\bibfnamefont{P.}~\bibnamefont{{Qian}}}, \bibnamefont{and}
  \bibinfo{author}{\bibfnamefont{C.~N.} \bibnamefont{{Hunter}}},
  \bibinfo{journal}{Chem. Phys. Lett.} \textbf{\bibinfo{volume}{414}},
  \bibinfo{pages}{359} (\bibinfo{year}{2005}).

\bibitem[{\citenamefont{{Cheng} and {Silbey}}(2006)}]{silbey_PRL_2006}
\bibinfo{author}{\bibfnamefont{Y.~C.} \bibnamefont{{Cheng}}} \bibnamefont{and}
  \bibinfo{author}{\bibfnamefont{R.~J.} \bibnamefont{{Silbey}}},
  \bibinfo{journal}{Phys. Rev. Lett.} \textbf{\bibinfo{volume}{96}},
  \bibinfo{pages}{028103} (\bibinfo{year}{2006}).

\bibitem[{\citenamefont{{Olaya-Castro}
  et~al.}(2008)\citenamefont{{Olaya-Castro}, {Lee}, {Fassioli Olsen}, and
  {Johnson}}}]{OlayaLF+08}
\bibinfo{author}{\bibfnamefont{A.}~\bibnamefont{{Olaya-Castro}}},
  \bibinfo{author}{\bibfnamefont{C.~F.} \bibnamefont{{Lee}}},
  \bibinfo{author}{\bibfnamefont{F.}~\bibnamefont{{Fassioli Olsen}}},
  \bibnamefont{and} \bibinfo{author}{\bibfnamefont{N.~F.}
  \bibnamefont{{Johnson}}}, \bibinfo{journal}{Phys. Rev. B}
  \textbf{\bibinfo{volume}{78}}, \bibinfo{pages}{085115}
  (\bibinfo{year}{2008}).

\bibitem[{\citenamefont{{Engel} et~al.}(2007)\citenamefont{{Engel}, {Calhoun},
  {Read}, {Ahn}, {Mancal}, {Cheng}, {Blankenship}, and
  {Fleming}}}]{Engel_Nature2006}
\bibinfo{author}{\bibfnamefont{G.~S.} \bibnamefont{{Engel}}},
  \bibinfo{author}{\bibfnamefont{T.~R.} \bibnamefont{{Calhoun}}},
  \bibinfo{author}{\bibfnamefont{E.~L.} \bibnamefont{{Read}}},
  \bibinfo{author}{\bibfnamefont{T.-K.} \bibnamefont{{Ahn}}},
  \bibinfo{author}{\bibfnamefont{T.}~\bibnamefont{{Mancal}}},
  \bibinfo{author}{\bibfnamefont{Y.~C.} \bibnamefont{{Cheng}}},
  \bibinfo{author}{\bibfnamefont{R.~E.} \bibnamefont{{Blankenship}}},
  \bibnamefont{and} \bibinfo{author}{\bibfnamefont{G.~R.}
  \bibnamefont{{Fleming}}}, \bibinfo{journal}{Nature}
  \textbf{\bibinfo{volume}{446}}, \bibinfo{pages}{782} (\bibinfo{year}{2007}).

\bibitem[{\citenamefont{{Hayes} et~al.}(2010)\citenamefont{{Hayes},
  {Panitchayangkoon}, {Fransted}, {Caram}, {Wen}, {Freed}, and
  {Engel}}}]{EngelNJP}
\bibinfo{author}{\bibfnamefont{D.}~\bibnamefont{{Hayes}}},
  \bibinfo{author}{\bibfnamefont{G.}~\bibnamefont{{Panitchayangkoon}}},
  \bibinfo{author}{\bibfnamefont{K.~A.} \bibnamefont{{Fransted}}},
  \bibinfo{author}{\bibfnamefont{J.~R.} \bibnamefont{{Caram}}},
  \bibinfo{author}{\bibfnamefont{J.}~\bibnamefont{{Wen}}},
  \bibinfo{author}{\bibfnamefont{K.~F.} \bibnamefont{{Freed}}},
  \bibnamefont{and} \bibinfo{author}{\bibfnamefont{G.}~\bibnamefont{{Engel}}},
  \bibinfo{journal}{New J. Phys.} \textbf{\bibinfo{volume}{12}},
  \bibinfo{pages}{065042} (\bibinfo{year}{2010}).

\bibitem[{\citenamefont{{Collini} et~al.}(2010)\citenamefont{{Collini}, {Wong},
  {Wilk}, {Curmi}, {Brumer}, and {Scholes}}}]{ColliniWW+10}
\bibinfo{author}{\bibfnamefont{E.}~\bibnamefont{{Collini}}},
  \bibinfo{author}{\bibfnamefont{C.~Y.} \bibnamefont{{Wong}}},
  \bibinfo{author}{\bibfnamefont{K.~E.} \bibnamefont{{Wilk}}},
  \bibinfo{author}{\bibfnamefont{P.~M.~G.} \bibnamefont{{Curmi}}},
  \bibinfo{author}{\bibfnamefont{P.}~\bibnamefont{{Brumer}}}, \bibnamefont{and}
  \bibinfo{author}{\bibfnamefont{G.~D.} \bibnamefont{{Scholes}}},
  \bibinfo{journal}{Nature} \textbf{\bibinfo{volume}{463}},
  \bibinfo{pages}{644} (\bibinfo{year}{2010}).

\bibitem[{\citenamefont{{Panitchayangkoon}
  et~al.}(2010)\citenamefont{{Panitchayangkoon}, {Hayes}, {Fransted}, {Caram},
  {Harel}, {Wen}, {Blankenship}, and {Engel}}}]{PanitchayangkoonVA+10}
\bibinfo{author}{\bibfnamefont{G.}~\bibnamefont{{Panitchayangkoon}}},
  \bibinfo{author}{\bibfnamefont{D.}~\bibnamefont{{Hayes}}},
  \bibinfo{author}{\bibfnamefont{K.~A.} \bibnamefont{{Fransted}}},
  \bibinfo{author}{\bibfnamefont{J.~R.} \bibnamefont{{Caram}}},
  \bibinfo{author}{\bibfnamefont{E.}~\bibnamefont{{Harel}}},
  \bibinfo{author}{\bibfnamefont{J.}~\bibnamefont{{Wen}}},
  \bibinfo{author}{\bibfnamefont{R.~E.} \bibnamefont{{Blankenship}}},
  \bibnamefont{and} \bibinfo{author}{\bibfnamefont{G.~S.}
  \bibnamefont{{Engel}}}, \bibinfo{journal}{Proc. Natl. Acad. Sci. USA}
  \textbf{\bibinfo{volume}{107}}, \bibinfo{pages}{12766}
  (\bibinfo{year}{2010}).

\bibitem[{\citenamefont{{Hildner} et~al.}(2013)\citenamefont{{Hildner},
  {Brinks}, {Nieder}, {Cogdell}, and {van Hulst}}}]{Hildner2013}
\bibinfo{author}{\bibfnamefont{R.}~\bibnamefont{{Hildner}}},
  \bibinfo{author}{\bibfnamefont{D.}~\bibnamefont{{Brinks}}},
  \bibinfo{author}{\bibfnamefont{J.~B.} \bibnamefont{{Nieder}}},
  \bibinfo{author}{\bibfnamefont{R.~J.} \bibnamefont{{Cogdell}}},
  \bibnamefont{and} \bibinfo{author}{\bibfnamefont{N.~F.} \bibnamefont{{van
  Hulst}}}, \bibinfo{journal}{Science} \textbf{\bibinfo{volume}{340}},
  \bibinfo{pages}{1448} (\bibinfo{year}{2013}).

\bibitem[{\citenamefont{{Romero} et~al.}(2014)\citenamefont{{Romero},
  {Augulis}, {Novoderezhkin}, {Ferretti}, {Thieme}, {Zigmantas}, and {van
  Grondelle}}}]{Romero_NPhys2014}
\bibinfo{author}{\bibfnamefont{E.}~\bibnamefont{{Romero}}},
  \bibinfo{author}{\bibfnamefont{R.}~\bibnamefont{{Augulis}}},
  \bibinfo{author}{\bibfnamefont{V.~I.} \bibnamefont{{Novoderezhkin}}},
  \bibinfo{author}{\bibfnamefont{M.}~\bibnamefont{{Ferretti}}},
  \bibinfo{author}{\bibfnamefont{J.}~\bibnamefont{{Thieme}}},
  \bibinfo{author}{\bibfnamefont{D.}~\bibnamefont{{Zigmantas}}},
  \bibnamefont{and} \bibinfo{author}{\bibfnamefont{R.}~\bibnamefont{{van
  Grondelle}}}, \bibinfo{journal}{Nature Phys.} \textbf{\bibinfo{volume}{10}},
  \bibinfo{pages}{676} (\bibinfo{year}{2014}).

\bibitem[{\citenamefont{{Fuller} et~al.}(2014)\citenamefont{{Fuller}, {Pan},
  {Gelzinis}, {Butkus}, {Senlik}, {Wilcox}, {Yocum}, {Valkunas},
  {Abramavicius}, and {Ogilvie}}}]{Ogilvie_NChem2014}
\bibinfo{author}{\bibfnamefont{F.~D.} \bibnamefont{{Fuller}}},
  \bibinfo{author}{\bibfnamefont{J.}~\bibnamefont{{Pan}}},
  \bibinfo{author}{\bibfnamefont{A.}~\bibnamefont{{Gelzinis}}},
  \bibinfo{author}{\bibfnamefont{V.}~\bibnamefont{{Butkus}}},
  \bibinfo{author}{\bibfnamefont{S.~S.} \bibnamefont{{Senlik}}},
  \bibinfo{author}{\bibfnamefont{D.~E.} \bibnamefont{{Wilcox}}},
  \bibinfo{author}{\bibfnamefont{C.~F.} \bibnamefont{{Yocum}}},
  \bibinfo{author}{\bibfnamefont{L.}~\bibnamefont{{Valkunas}}},
  \bibinfo{author}{\bibfnamefont{D.}~\bibnamefont{{Abramavicius}}},
  \bibnamefont{and} \bibinfo{author}{\bibfnamefont{J.~P.}
  \bibnamefont{{Ogilvie}}}, \bibinfo{journal}{Nature Chem.}
  \textbf{\bibinfo{volume}{6}}, \bibinfo{pages}{706} (\bibinfo{year}{2014}).

\bibitem[{\citenamefont{{Tiwari} et~al.}(2013)\citenamefont{{Tiwari}, {Peters},
  and {Jonas}}}]{Jonas_PNAS2012}
\bibinfo{author}{\bibfnamefont{V.}~\bibnamefont{{Tiwari}}},
  \bibinfo{author}{\bibfnamefont{W.~K.} \bibnamefont{{Peters}}},
  \bibnamefont{and} \bibinfo{author}{\bibfnamefont{D.~M.}
  \bibnamefont{{Jonas}}}, \bibinfo{journal}{Proc. Natl Acad. Sci. USA}
  \textbf{\bibinfo{volume}{110}}, \bibinfo{pages}{1203} (\bibinfo{year}{2013}).

\bibitem[{\citenamefont{{Plenio} et~al.}(2013)\citenamefont{{Plenio},
  {Almeida}, and {Huelga}}}]{Plenio_JCP2013}
\bibinfo{author}{\bibfnamefont{M.~B.} \bibnamefont{{Plenio}}},
  \bibinfo{author}{\bibfnamefont{J.}~\bibnamefont{{Almeida}}},
  \bibnamefont{and} \bibinfo{author}{\bibfnamefont{S.~F.}
  \bibnamefont{{Huelga}}}, \bibinfo{journal}{J. Chem. Phys.}
  \textbf{\bibinfo{volume}{139}}, \bibinfo{pages}{235102}
  (\bibinfo{year}{2013}).

\bibitem[{\citenamefont{{Caycedo-Soler}
  et~al.}(2012)\citenamefont{{Caycedo-Soler}, {Chin}, {Almeida}, {Huelga}, and
  {Plenio}}}]{Caycedo_2012}
\bibinfo{author}{\bibfnamefont{F.}~\bibnamefont{{Caycedo-Soler}}},
  \bibinfo{author}{\bibfnamefont{A.~W.} \bibnamefont{{Chin}}},
  \bibinfo{author}{\bibfnamefont{J.}~\bibnamefont{{Almeida}}},
  \bibinfo{author}{\bibfnamefont{S.~F.} \bibnamefont{{Huelga}}},
  \bibnamefont{and} \bibinfo{author}{\bibfnamefont{M.~B.}
  \bibnamefont{{Plenio}}}, \bibinfo{journal}{J. Chem. Phys.}
  \textbf{\bibinfo{volume}{136}}, \bibinfo{pages}{155102}
  (\bibinfo{year}{2012}).

\bibitem[{\citenamefont{{Christensson}
  et~al.}(2012)\citenamefont{{Christensson}, {Kauffmann}, {Pullerits}, and
  {Man{\v c}al}}}]{Christensson_JPCB2012}
\bibinfo{author}{\bibfnamefont{N.}~\bibnamefont{{Christensson}}},
  \bibinfo{author}{\bibfnamefont{H.~F.} \bibnamefont{{Kauffmann}}},
  \bibinfo{author}{\bibfnamefont{T.}~\bibnamefont{{Pullerits}}},
  \bibnamefont{and} \bibinfo{author}{\bibfnamefont{T.}~\bibnamefont{{Man{\v
  c}al}}}, \bibinfo{journal}{J. Phys. Chem. B} \textbf{\bibinfo{volume}{116}},
  \bibinfo{pages}{7449} (\bibinfo{year}{2012}).

\bibitem[{\citenamefont{{Kolli} et~al.}(2012)\citenamefont{{Kolli}, {O'Reilly},
  {Scholes}, and {Olaya-Castro}}}]{KolliRS+12}
\bibinfo{author}{\bibfnamefont{A.}~\bibnamefont{{Kolli}}},
  \bibinfo{author}{\bibfnamefont{E.~J.} \bibnamefont{{O'Reilly}}},
  \bibinfo{author}{\bibfnamefont{G.~D.} \bibnamefont{{Scholes}}},
  \bibnamefont{and}
  \bibinfo{author}{\bibfnamefont{A.}~\bibnamefont{{Olaya-Castro}}},
  \bibinfo{journal}{J. Chem. Phys.} \textbf{\bibinfo{volume}{137}},
  \bibinfo{pages}{174109} (\bibinfo{year}{2012}).

\bibitem[{\citenamefont{{Chin} et~al.}(2013)\citenamefont{{Chin}, {Prior},
  {Rosenbach}, {Caycedo-Soler}, {Huelga}, and {Plenio}}}]{ChinPR+13}
\bibinfo{author}{\bibfnamefont{A.~W.} \bibnamefont{{Chin}}},
  \bibinfo{author}{\bibfnamefont{J.}~\bibnamefont{{Prior}}},
  \bibinfo{author}{\bibfnamefont{R.}~\bibnamefont{{Rosenbach}}},
  \bibinfo{author}{\bibfnamefont{F.}~\bibnamefont{{Caycedo-Soler}}},
  \bibinfo{author}{\bibfnamefont{S.~F.} \bibnamefont{{Huelga}}},
  \bibnamefont{and} \bibinfo{author}{\bibfnamefont{M.~B.}
  \bibnamefont{{Plenio}}}, \bibinfo{journal}{Nature Phys.}
  \textbf{\bibinfo{volume}{9}}, \bibinfo{pages}{113} (\bibinfo{year}{2013}).

\bibitem[{\citenamefont{{Huelga} and {Plenio}}(2013)}]{HuelgaP13}
\bibinfo{author}{\bibfnamefont{S.~F.} \bibnamefont{{Huelga}}} \bibnamefont{and}
  \bibinfo{author}{\bibfnamefont{M.~B.} \bibnamefont{{Plenio}}},
  \bibinfo{journal}{Contemporary Physics} \textbf{\bibinfo{volume}{54}},
  \bibinfo{pages}{181} (\bibinfo{year}{2013}).

\bibitem[{\citenamefont{{Chin} et~al.}(2010)\citenamefont{{Chin}, {Datta},
  {Carusso}, {Huelga}, and {Plenio}}}]{Chin2010}
\bibinfo{author}{\bibfnamefont{A.}~\bibnamefont{{Chin}}},
  \bibinfo{author}{\bibfnamefont{A.}~\bibnamefont{{Datta}}},
  \bibinfo{author}{\bibfnamefont{F.}~\bibnamefont{{Carusso}}},
  \bibinfo{author}{\bibfnamefont{S.~F.} \bibnamefont{{Huelga}}},
  \bibnamefont{and} \bibinfo{author}{\bibfnamefont{M.~B.}
  \bibnamefont{{Plenio}}}, \bibinfo{journal}{New J. Phys.}
  \textbf{\bibinfo{volume}{12}}, \bibinfo{pages}{065002}
  (\bibinfo{year}{2010}).

\bibitem[{\citenamefont{{Mohseni} et~al.}(2008)\citenamefont{{Mohseni},
  {Rebentrost}, {Lloyd}, and {Aspuru-Guzik}}}]{Mohseni08}
\bibinfo{author}{\bibfnamefont{M.}~\bibnamefont{{Mohseni}}},
  \bibinfo{author}{\bibfnamefont{P.}~\bibnamefont{{Rebentrost}}},
  \bibinfo{author}{\bibfnamefont{S.}~\bibnamefont{{Lloyd}}}, \bibnamefont{and}
  \bibinfo{author}{\bibfnamefont{A.}~\bibnamefont{{Aspuru-Guzik}}},
  \bibinfo{journal}{J. Chem. Phys.} \textbf{\bibinfo{volume}{129}},
  \bibinfo{pages}{174106} (\bibinfo{year}{2008}).

\bibitem[{\citenamefont{{Plenio} and {Huelga}}(2008)}]{PlenioH08}
\bibinfo{author}{\bibfnamefont{M.~B.} \bibnamefont{{Plenio}}} \bibnamefont{and}
  \bibinfo{author}{\bibfnamefont{S.~F.} \bibnamefont{{Huelga}}},
  \bibinfo{journal}{New J. Phys.} \textbf{\bibinfo{volume}{10}},
  \bibinfo{pages}{113019} (\bibinfo{year}{2008}).

\bibitem[{\citenamefont{{Womick} and {Moran}}(2011)}]{Womick_2011}
\bibinfo{author}{\bibfnamefont{J.}~\bibnamefont{{Womick}}} \bibnamefont{and}
  \bibinfo{author}{\bibfnamefont{A.}~\bibnamefont{{Moran}}},
  \bibinfo{journal}{J. Phys. Chem. B} \textbf{\bibinfo{volume}{115}},
  \bibinfo{pages}{1347} (\bibinfo{year}{2011}).

\bibitem[{\citenamefont{{Str\"umpfer} and
  {Schulten}}(2012)}]{Strumpfer_JCP2012}
\bibinfo{author}{\bibfnamefont{J.}~\bibnamefont{{Str\"umpfer}}}
  \bibnamefont{and}
  \bibinfo{author}{\bibfnamefont{K.}~\bibnamefont{{Schulten}}},
  \bibinfo{journal}{J. Chem. Phys.} \textbf{\bibinfo{volume}{137}},
  \bibinfo{pages}{065101} (\bibinfo{year}{2012}).

\bibitem[{\citenamefont{{Stahlberg} et~al.}(1998)\citenamefont{{Stahlberg},
  {Dubochet}, {Vogel}, and {Ghosh}}}]{Ghosh}
\bibinfo{author}{\bibfnamefont{H.}~\bibnamefont{{Stahlberg}}},
  \bibinfo{author}{\bibfnamefont{J.}~\bibnamefont{{Dubochet}}},
  \bibinfo{author}{\bibfnamefont{H.}~\bibnamefont{{Vogel}}}, \bibnamefont{and}
  \bibinfo{author}{\bibfnamefont{R.}~\bibnamefont{{Ghosh}}},
  \bibinfo{journal}{J. Mol. Biol.} \textbf{\bibinfo{volume}{282}},
  \bibinfo{pages}{819} (\bibinfo{year}{1998}).

\bibitem[{\citenamefont{{Stomp} et~al.}(2007)\citenamefont{{Stomp}, {Huisman},
  {Stal}, and {Matthijs}}}]{Stomp_2007ISME}
\bibinfo{author}{\bibfnamefont{M.}~\bibnamefont{{Stomp}}},
  \bibinfo{author}{\bibfnamefont{J.}~\bibnamefont{{Huisman}}},
  \bibinfo{author}{\bibfnamefont{L.~J.} \bibnamefont{{Stal}}},
  \bibnamefont{and} \bibinfo{author}{\bibfnamefont{H.~C.~P.}
  \bibnamefont{{Matthijs}}}, \bibinfo{journal}{ISME J.}
  \textbf{\bibinfo{volume}{1}}, \bibinfo{pages}{271} (\bibinfo{year}{2007}).

\bibitem[{\citenamefont{{Jimenez} et~al.}(1997)\citenamefont{{Jimenez}, {van
  Mourik}, {Young Yu}, and {Flemming}}}]{Jimenez_JPCB1997}
\bibinfo{author}{\bibfnamefont{R.}~\bibnamefont{{Jimenez}}},
  \bibinfo{author}{\bibfnamefont{F.}~\bibnamefont{{van Mourik}}},
  \bibinfo{author}{\bibfnamefont{J.}~\bibnamefont{{Young Yu}}},
  \bibnamefont{and} \bibinfo{author}{\bibfnamefont{G.~R.}
  \bibnamefont{{Flemming}}}, \bibinfo{journal}{J. Phys. Chem. B}
  \textbf{\bibinfo{volume}{101}}, \bibinfo{pages}{7350} (\bibinfo{year}{1997}).

\bibitem[{\citenamefont{{ Karrasch} et~al.}(1995)\citenamefont{{ Karrasch},
  {Bullough}, and {Ghosh}}}]{karrasch95}
\bibinfo{author}{\bibfnamefont{S.}~\bibnamefont{{ Karrasch}}},
  \bibinfo{author}{\bibfnamefont{P.~A.} \bibnamefont{{Bullough}}},
  \bibnamefont{and} \bibinfo{author}{\bibfnamefont{R.}~\bibnamefont{{Ghosh}}},
  \bibinfo{journal}{EMBO J.} \textbf{\bibinfo{volume}{14}},
  \bibinfo{pages}{631} (\bibinfo{year}{1995}).

\bibitem[{\citenamefont{{Roszak} et~al.}(2003)\citenamefont{{Roszak}, {Howard},
  {Southall}, {Gardiner}, {Law}, {Isaacs}, and
  {Cogdell}}}]{Cogdell_Science_2003}
\bibinfo{author}{\bibfnamefont{A.~W.} \bibnamefont{{Roszak}}},
  \bibinfo{author}{\bibfnamefont{T.~D.} \bibnamefont{{Howard}}},
  \bibinfo{author}{\bibfnamefont{J.}~\bibnamefont{{Southall}}},
  \bibinfo{author}{\bibfnamefont{A.~T.} \bibnamefont{{Gardiner}}},
  \bibinfo{author}{\bibfnamefont{C.~J.} \bibnamefont{{Law}}},
  \bibinfo{author}{\bibfnamefont{N.~W.} \bibnamefont{{Isaacs}}},
  \bibnamefont{and} \bibinfo{author}{\bibfnamefont{R.~J.}
  \bibnamefont{{Cogdell}}}, \bibinfo{journal}{Science}
  \textbf{\bibinfo{volume}{302}}, \bibinfo{pages}{1969} (\bibinfo{year}{2003}).

\bibitem[{\citenamefont{{Monshouwer} et~al.}(1997)\citenamefont{{Monshouwer},
  {Abrahamsson}, {van Mourik}, and {van Grondelle}}}]{vanGrondelle_1997JPCB}
\bibinfo{author}{\bibfnamefont{R.}~\bibnamefont{{Monshouwer}}},
  \bibinfo{author}{\bibfnamefont{M.}~\bibnamefont{{Abrahamsson}}},
  \bibinfo{author}{\bibfnamefont{F.}~\bibnamefont{{van Mourik}}},
  \bibnamefont{and} \bibinfo{author}{\bibfnamefont{R.}~\bibnamefont{{van
  Grondelle}}}, \bibinfo{journal}{J. Phys. Chem. B}
  \textbf{\bibinfo{volume}{101}}, \bibinfo{pages}{7241} (\bibinfo{year}{1997}).

\bibitem[{\citenamefont{{Ghosh} et~al.}(1988)\citenamefont{{Ghosh}, {Hauser},
  and {Bachofen}}}]{Ghosh_1988BioC}
\bibinfo{author}{\bibfnamefont{R.}~\bibnamefont{{Ghosh}}},
  \bibinfo{author}{\bibfnamefont{H.}~\bibnamefont{{Hauser}}}, \bibnamefont{and}
  \bibinfo{author}{\bibfnamefont{R.}~\bibnamefont{{Bachofen}}},
  \bibinfo{journal}{Biochemistry} \textbf{\bibinfo{volume}{27}},
  \bibinfo{pages}{1004} (\bibinfo{year}{1988}).

\bibitem[{\citenamefont{{Zureck}}(2003)}]{Zureck_RMP2003}
\bibinfo{author}{\bibfnamefont{W.~H.} \bibnamefont{{Zureck}}},
  \bibinfo{journal}{Rev. Mod. Phys.} \textbf{\bibinfo{volume}{75}},
  \bibinfo{pages}{715} (\bibinfo{year}{2003}).

\end{thebibliography}

\newpage
\begin{thebibliography}{22}
\expandafter\ifx\csname natexlab\endcsname\relax\def\natexlab#1{#1}\fi
\expandafter\ifx\csname bibnamefont\endcsname\relax
  \def\bibnamefont#1{#1}\fi
\expandafter\ifx\csname bibfnamefont\endcsname\relax
  \def\bibfnamefont#1{#1}\fi
\expandafter\ifx\csname citenamefont\endcsname\relax
  \def\citenamefont#1{#1}\fi
\expandafter\ifx\csname url\endcsname\relax
  \def\url#1{\texttt{#1}}\fi
\expandafter\ifx\csname urlprefix\endcsname\relax\def\urlprefix{URL }\fi
\providecommand{\bibinfo}[2]{#2}
\providecommand{\eprint}[2][]{\url{#2}}

\bibitem[{\citenamefont{{Gerken}
  et~al.}(2003{\natexlab{a}})\citenamefont{{Gerken}, {Jelezko}, {G\"otze},
  {Bransch\"adel}, {Tietz}, {Ghosh}, and {Wrachtrup}}}]{Gerken_JPC2003}
\bibinfo{author}{\bibfnamefont{U.}~\bibnamefont{{Gerken}}},
  \bibinfo{author}{\bibfnamefont{F.}~\bibnamefont{{Jelezko}}},
  \bibinfo{author}{\bibfnamefont{B.}~\bibnamefont{{G\"otze}}},
  \bibinfo{author}{\bibfnamefont{M.}~\bibnamefont{{Bransch\"adel}}},
  \bibinfo{author}{\bibfnamefont{C.}~\bibnamefont{{Tietz}}},
  \bibinfo{author}{\bibfnamefont{R.}~\bibnamefont{{Ghosh}}}, \bibnamefont{and}
  \bibinfo{author}{\bibfnamefont{J.~.} \bibnamefont{{Wrachtrup}}},
  \bibinfo{journal}{J. Phys. Chem. B} \textbf{\bibinfo{volume}{107}},
  \bibinfo{pages}{338} (\bibinfo{year}{2003}{\natexlab{a}}).

\bibitem[{\citenamefont{{Gerken}
  et~al.}(2003{\natexlab{b}})\citenamefont{{Gerken}, {Lupo}, {Tietz},
  {Wrachtrup}, and {Ghosh}}}]{Ghosh_Biochem_2003}
\bibinfo{author}{\bibfnamefont{U.}~\bibnamefont{{Gerken}}},
  \bibinfo{author}{\bibfnamefont{D.}~\bibnamefont{{Lupo}}},
  \bibinfo{author}{\bibfnamefont{C.}~\bibnamefont{{Tietz}}},
  \bibinfo{author}{\bibfnamefont{J.}~\bibnamefont{{Wrachtrup}}},
  \bibnamefont{and} \bibinfo{author}{\bibfnamefont{R.}~\bibnamefont{{Ghosh}}},
  \bibinfo{journal}{Biochemistry} \textbf{\bibinfo{volume}{42}},
  \bibinfo{pages}{10354} (\bibinfo{year}{2003}{\natexlab{b}}).

\bibitem[{\citenamefont{{Niwa} et~al.}(2014)\citenamefont{{Niwa}, {Yu},
  {Takeda}, {Hirano}, {Kawakami}, {Wang-Otomo}, and {Miki}}}]{Niway_2014Nat}
\bibinfo{author}{\bibfnamefont{S.}~\bibnamefont{{Niwa}}},
  \bibinfo{author}{\bibfnamefont{L.~J.} \bibnamefont{{Yu}}},
  \bibinfo{author}{\bibfnamefont{K.}~\bibnamefont{{Takeda}}},
  \bibinfo{author}{\bibfnamefont{Y.}~\bibnamefont{{Hirano}}},
  \bibinfo{author}{\bibfnamefont{T.}~\bibnamefont{{Kawakami}}},
  \bibinfo{author}{\bibfnamefont{Z.~Y.} \bibnamefont{{Wang-Otomo}}},
  \bibnamefont{and} \bibinfo{author}{\bibfnamefont{K.}~\bibnamefont{{Miki}}},
  \bibinfo{journal}{Nature} \textbf{\bibinfo{volume}{508}},
  \bibinfo{pages}{228} (\bibinfo{year}{2014}).

\bibitem[{\citenamefont{{Weiss, Jr.}}(1972)}]{Weiss72}
\bibinfo{author}{\bibfnamefont{C.}~\bibnamefont{{Weiss, Jr.}}},
  \bibinfo{journal}{J. Mol. Spectrosc.} \textbf{\bibinfo{volume}{44}},
  \bibinfo{pages}{37} (\bibinfo{year}{1972}).

\bibitem[{\citenamefont{{Autenrieth}}(2002)}]{Autenrieth_2002}
\bibinfo{author}{\bibfnamefont{F.}~\bibnamefont{{Autenrieth}}},
  \bibinfo{type}{Diploma thesis}, \bibinfo{school}{University of Stuttgart}
  (\bibinfo{year}{2002}).

\bibitem[{\citenamefont{{Jimenez} et~al.}(1997)\citenamefont{{Jimenez}, {van
  Mourik}, {Young Yu}, and {Flemming}}}]{Jimenez_JPCB1997}
\bibinfo{author}{\bibfnamefont{R.}~\bibnamefont{{Jimenez}}},
  \bibinfo{author}{\bibfnamefont{F.}~\bibnamefont{{van Mourik}}},
  \bibinfo{author}{\bibfnamefont{J.}~\bibnamefont{{Young Yu}}},
  \bibnamefont{and} \bibinfo{author}{\bibfnamefont{G.~R.}
  \bibnamefont{{Flemming}}}, \bibinfo{journal}{J. Phys. Chem. B}
  \textbf{\bibinfo{volume}{101}}, \bibinfo{pages}{7350} (\bibinfo{year}{1997}).

\bibitem[{\citenamefont{Jordanides et~al.}(2001)\citenamefont{Jordanides,
  Scholes, and Fleming}}]{Jordanides_2001JPCB}
\bibinfo{author}{\bibfnamefont{X.~J.} \bibnamefont{Jordanides}},
  \bibinfo{author}{\bibfnamefont{G.~D.} \bibnamefont{Scholes}},
  \bibnamefont{and} \bibinfo{author}{\bibfnamefont{G.~R.}
  \bibnamefont{Fleming}}, \bibinfo{journal}{J. Phys. Chem. B}
  \textbf{\bibinfo{volume}{105}}, \bibinfo{pages}{1652} (\bibinfo{year}{2001}).

\bibitem[{\citenamefont{{Monshouwer} et~al.}(1997)\citenamefont{{Monshouwer},
  {Abrahamsson}, {van Mourik}, and {van Grondelle}}}]{vanGrondelle_1997JPCB}
\bibinfo{author}{\bibfnamefont{R.}~\bibnamefont{{Monshouwer}}},
  \bibinfo{author}{\bibfnamefont{M.}~\bibnamefont{{Abrahamsson}}},
  \bibinfo{author}{\bibfnamefont{F.}~\bibnamefont{{van Mourik}}},
  \bibnamefont{and} \bibinfo{author}{\bibfnamefont{R.}~\bibnamefont{{van
  Grondelle}}}, \bibinfo{journal}{J. Phys. Chem. B}
  \textbf{\bibinfo{volume}{101}}, \bibinfo{pages}{7241} (\bibinfo{year}{1997}).

\bibitem[{\citenamefont{{Georgakopoulou}
  et~al.}(2006)\citenamefont{{Georgakopoulou}, {van Grondelle}, and {van der
  Zwan}}}]{VanGrondelle_JP2006}
\bibinfo{author}{\bibfnamefont{S.}~\bibnamefont{{Georgakopoulou}}},
  \bibinfo{author}{\bibfnamefont{R.}~\bibnamefont{{van Grondelle}}},
  \bibnamefont{and} \bibinfo{author}{\bibfnamefont{G.}~\bibnamefont{{van der
  Zwan}}}, \bibinfo{journal}{J. Phys. Chem. B} \textbf{\bibinfo{volume}{110}},
  \bibinfo{pages}{3344} (\bibinfo{year}{2006}).

\bibitem[{\citenamefont{{Freiberg} and {Timpmann}}(1992)}]{Freiberg92}
\bibinfo{author}{\bibfnamefont{A.}~\bibnamefont{{Freiberg}}} \bibnamefont{and}
  \bibinfo{author}{\bibfnamefont{K.}~\bibnamefont{{Timpmann}}},
  \bibinfo{journal}{Photochem. Photobiol. B} \textbf{\bibinfo{volume}{15}},
  \bibinfo{pages}{151} (\bibinfo{year}{1992}).

\bibitem[{\citenamefont{{Kubo}}(1962)}]{Kubo_1962JPSJ}
\bibinfo{author}{\bibfnamefont{R.}~\bibnamefont{{Kubo}}}, \bibinfo{journal}{J.
  Phys. Soc. Jpn.} \textbf{\bibinfo{volume}{17}}, \bibinfo{pages}{1100}
  (\bibinfo{year}{1962}).

\bibitem[{\citenamefont{{Groot} et~al.}(1998)\citenamefont{{Groot}, {Yu},
  {Agarwal}, {Norris}, and {Fleming}}}]{Flemming_1998JPCB}
\bibinfo{author}{\bibfnamefont{M.~L.} \bibnamefont{{Groot}}},
  \bibinfo{author}{\bibfnamefont{J.~Y.} \bibnamefont{{Yu}}},
  \bibinfo{author}{\bibfnamefont{R.}~\bibnamefont{{Agarwal}}},
  \bibinfo{author}{\bibfnamefont{J.~N.} \bibnamefont{{Norris}}},
  \bibnamefont{and} \bibinfo{author}{\bibfnamefont{G.~R.}
  \bibnamefont{{Fleming}}}, \bibinfo{journal}{J. Phys. Chem. B}
  \textbf{\bibinfo{volume}{102}}, \bibinfo{pages}{5923} (\bibinfo{year}{1998}).

\bibitem[{\citenamefont{{Adolphs} and {Renger}}(2006)}]{Renger2006}
\bibinfo{author}{\bibfnamefont{J.}~\bibnamefont{{Adolphs}}} \bibnamefont{and}
  \bibinfo{author}{\bibfnamefont{T.}~\bibnamefont{{Renger}}},
  \bibinfo{journal}{Biophys. J.} \textbf{\bibinfo{volume}{91}},
  \bibinfo{pages}{2778} (\bibinfo{year}{2006}).

\bibitem[{\citenamefont{{Cohen Stuart} et~al.}(2011)\citenamefont{{Cohen
  Stuart}, {Vengris}, {Novoderezhkin}, {Cogdell}, {Hunter}, and {van
  Grondelle}}}]{Cohen_2011BioJ}
\bibinfo{author}{\bibfnamefont{T.~A.} \bibnamefont{{Cohen Stuart}}},
  \bibinfo{author}{\bibfnamefont{M.}~\bibnamefont{{Vengris}}},
  \bibinfo{author}{\bibfnamefont{V.~I.} \bibnamefont{{Novoderezhkin}}},
  \bibinfo{author}{\bibfnamefont{R.~J.} \bibnamefont{{Cogdell}}},
  \bibinfo{author}{\bibfnamefont{C.~N.} \bibnamefont{{Hunter}}},
  \bibnamefont{and} \bibinfo{author}{\bibfnamefont{R.}~\bibnamefont{{van
  Grondelle}}}, \bibinfo{journal}{Biophys. J.} \textbf{\bibinfo{volume}{100}},
  \bibinfo{pages}{2226} (\bibinfo{year}{2011}).

\bibitem[{\citenamefont{{Timpmann} et~al.}(2005)\citenamefont{{Timpmann},
  {Trinkunas}, {Qian}, and {Hunter}}}]{Timpmann_2005CPL}
\bibinfo{author}{\bibfnamefont{K.}~\bibnamefont{{Timpmann}}},
  \bibinfo{author}{\bibfnamefont{G.}~\bibnamefont{{Trinkunas}}},
  \bibinfo{author}{\bibfnamefont{P.}~\bibnamefont{{Qian}}}, \bibnamefont{and}
  \bibinfo{author}{\bibfnamefont{C.~N.} \bibnamefont{{Hunter}}},
  \bibinfo{journal}{Chem. Phys. Lett.} \textbf{\bibinfo{volume}{414}},
  \bibinfo{pages}{359} (\bibinfo{year}{2005}).

\bibitem[{\citenamefont{{Plenio} et~al.}(2013)\citenamefont{{Plenio},
  {Almeida}, and {Huelga}}}]{Plenio_JCP2013}
\bibinfo{author}{\bibfnamefont{M.~B.} \bibnamefont{{Plenio}}},
  \bibinfo{author}{\bibfnamefont{J.}~\bibnamefont{{Almeida}}},
  \bibnamefont{and} \bibinfo{author}{\bibfnamefont{S.~F.}
  \bibnamefont{{Huelga}}}, \bibinfo{journal}{J. Chem. Phys.}
  \textbf{\bibinfo{volume}{139}}, \bibinfo{pages}{235102}
  (\bibinfo{year}{2013}).

\bibitem[{\citenamefont{{Ort} and {Yocum}}(1996)}]{Ort96}
\bibinfo{editor}{\bibfnamefont{D.~R.} \bibnamefont{{Ort}}} \bibnamefont{and}
  \bibinfo{editor}{\bibfnamefont{C.~F.} \bibnamefont{{Yocum}}}, eds.,
  \emph{\bibinfo{title}{{Oxygenic Photosynthesis: the light reactions}}}
  (\bibinfo{publisher}{Kluwer Academic Publishers}, \bibinfo{year}{1996}).

\bibitem[{\citenamefont{{Damjanovi\'c}
  et~al.}(2002)\citenamefont{{Damjanovi\'c}, {Vaswani}, {Fromme}, and
  {Fleming}}}]{Damjanovic_02}
\bibinfo{author}{\bibfnamefont{A.}~\bibnamefont{{Damjanovi\'c}}},
  \bibinfo{author}{\bibfnamefont{H.~M.} \bibnamefont{{Vaswani}}},
  \bibinfo{author}{\bibfnamefont{P.}~\bibnamefont{{Fromme}}}, \bibnamefont{and}
  \bibinfo{author}{\bibfnamefont{G.~R.} \bibnamefont{{Fleming}}},
  \bibinfo{journal}{J. Phys. Chem. B} \textbf{\bibinfo{volume}{106}},
  \bibinfo{pages}{10251} (\bibinfo{year}{2002}).

\bibitem[{\citenamefont{{Raszewksi} and {Renger}}(2008)}]{Raszewksi_08}
\bibinfo{author}{\bibfnamefont{G.}~\bibnamefont{{Raszewksi}}} \bibnamefont{and}
  \bibinfo{author}{\bibfnamefont{T.}~\bibnamefont{{Renger}}},
  \bibinfo{journal}{J. Am. Chem. Soc.} \textbf{\bibinfo{volume}{130}},
  \bibinfo{pages}{4431} (\bibinfo{year}{2008}).

\bibitem[{\citenamefont{{Raszewksi} et~al.}(2005)\citenamefont{{Raszewksi},
  {Saenger}, and {Renger}}}]{Raszewksi_05}
\bibinfo{author}{\bibfnamefont{G.}~\bibnamefont{{Raszewksi}}},
  \bibinfo{author}{\bibfnamefont{W.}~\bibnamefont{{Saenger}}},
  \bibnamefont{and} \bibinfo{author}{\bibfnamefont{T.}~\bibnamefont{{Renger}}},
  \bibinfo{journal}{Biophys. J.} \textbf{\bibinfo{volume}{88}},
  \bibinfo{pages}{986} (\bibinfo{year}{2005}).

\bibitem[{\citenamefont{{Shiwei} et~al.}(2007)\citenamefont{{Shiwei},
  {Dahlbom}, {Canfield}, {Hush}, {Kobayashi}, and {Reimers}}}]{Yin_07}
\bibinfo{author}{\bibfnamefont{Y.}~\bibnamefont{{Shiwei}}},
  \bibinfo{author}{\bibfnamefont{M.~G.} \bibnamefont{{Dahlbom}}},
  \bibinfo{author}{\bibfnamefont{P.~J.} \bibnamefont{{Canfield}}},
  \bibinfo{author}{\bibfnamefont{N.~S.} \bibnamefont{{Hush}}},
  \bibinfo{author}{\bibfnamefont{R.}~\bibnamefont{{Kobayashi}}},
  \bibnamefont{and} \bibinfo{author}{\bibfnamefont{J.~R.}
  \bibnamefont{{Reimers}}}, \bibinfo{journal}{J. Phys. Chem. B}
  \textbf{\bibinfo{volume}{111}}, \bibinfo{pages}{9923} (\bibinfo{year}{2007}).

\bibitem[{\citenamefont{{Konermann} and {Holzwarth}}(1996)}]{Holzwarth_96}
\bibinfo{author}{\bibfnamefont{L.}~\bibnamefont{{Konermann}}} \bibnamefont{and}
  \bibinfo{author}{\bibfnamefont{A.~R.} \bibnamefont{{Holzwarth}}},
  \bibinfo{journal}{Biochemistry} \textbf{\bibinfo{volume}{35}},
  \bibinfo{pages}{829} (\bibinfo{year}{1996}).

\end{thebibliography}
\end{document}